\documentclass[12pt]{article}
\pdfoutput=1

\usepackage[
 linktoc=all,
 colorlinks=true,
 linkcolor=blue,
 urlcolor=blue,
 citecolor=red
 ]{hyperref}

\usepackage[backend=bibtex,style=numeric-comp,sorting=none,maxbibnames=99, minbibnames=99]{biblatex}
\usepackage{mathtools,amssymb,mathrsfs,ytableau,microtype,tikz,pgfplots,dsfont}
\usetikzlibrary{decorations.pathmorphing}
\usetikzlibrary{decorations.markings}

\ytableausetup{centertableaux,boxsize=.5em}

\DeclareMathOperator{\sign}{sign}
\DeclareMathOperator{\rank}{rank}
\DeclareMathOperator{\tr}{tr}

\newcommand{\be}{\begin{equation}}
\newcommand{\ee}{\end{equation}}
\newcommand{\ba}{\begin{aligned}}
\newcommand{\ea}{\end{aligned}}
\newcommand{\bit}{\begin{itemize}}
\newcommand{\eit}{\end{itemize}}
\newcommand{\ben}{\begin{enumerate}}
\newcommand{\een}{\end{enumerate}}

\newcommand{\lb}{\left(}
\newcommand{\rb}{\right)}
\newcommand{\lbb}{\left[}
\newcommand{\rbb}{\right]}

\newcommand{\half}{\frac{1}{2}}

\newcommand{\cA}{\mathcal{A}}
\newcommand{\cB}{\mathcal{B}}
\newcommand{\cD}{\mathcal{D}}
\newcommand{\cJ}{\mathcal{J}}
\newcommand{\cR}{\mathcal{R}}
\newcommand{\cT}{\mathcal{T}}

\newcommand{\Z}{{\mathbb Z}}
\newcommand{\Q}{{\mathbb Q}}

\pgfplotsset{compat=1.18}

\usepackage[framemethod=TikZ]{mdframed}
\mdfsetup{%
skipabove=3pt,
skipbelow=2pt,
linecolor=black!0,
backgroundcolor=black!15,
roundcorner=3pt}

\renewcommand{\title}[1]{\vbox{\center\LARGE{#1}}\vspace{5mm}}
\renewcommand{\author}[1]{\vbox{\center\large#1}\vspace{5mm}}
\newcommand{\address}[1]{\vbox{\center\em#1}}

\makeatletter\@addtoreset{equation}{section}\makeatother

\usepackage{moresize}

\begin{document}

\begin{titlepage}
\begin{center}
\vskip-10pt
\title{\makebox[\textwidth]{\HUGE{
Monopoles, Scattering, and Generalized Symmetries
}}}
 \vspace{10mm}
 Marieke van Beest,${}^a\,$\footnote{\href{mailto:mvanbeest@scgp.stonybrook.edu}{\tt mvanbeest@scgp.stonybrook.edu}}
 Philip Boyle Smith,${}^b\,$\footnote{\href{mailto:philip.boyle.smith@ipmu.jp}{\tt philip.boyle.smith@ipmu.jp}}
 Diego Delmastro,${}^a\,$\footnote{\href{mailto:ddelmastro@scgp.stonybrook.edu}{\tt ddelmastro@scgp.stonybrook.edu}}
 Zohar Komargodski,${}^a\,$\footnote{\href{mailto:zkomargo@gmail.com}{\tt zkomargo@gmail.com}}
 David Tong.${}^c\,$\footnote{\href{mailto:d.tong@damtp.cam.ac.uk}{\tt d.tong@damtp.cam.ac.uk}}
 \vskip 7mm
 \address{
 ${}^a$ Simons Center for Geometry and Physics,\\
 SUNY, Stony Brook, NY 11794, USA}
 \address{
 ${}^b$ Kavli Institute for the Physics and Mathematics of the Universe (WPI),\\ University of Tokyo, Kashiwa, Chiba 277-8583, Japan}
  \address{
 ${}^c$ Department of Applied Mathematics and Theoretical Physics,\\ University of Cambridge, CB3 0WA, UK}
\end{center}

\abstract{We reconsider the  problem of electrically charged, massless fermions scattering off magnetic monopoles. The interpretation of the outgoing states has long been a puzzle as, in certain circumstances, they necessarily carry fractional quantum numbers. We argue that consistency requires such outgoing particles to be attached to a topological co-dimension 1 surface, which ends on the  monopole. This surface cannot participate in a 2-group with the magnetic 1-form symmetry and is often non-invertible. Equivalently, the outgoing radiation lies in a twisted sector and not in the original Fock space. The outgoing radiation therefore not only carries unconventional flavor quantum numbers, but is often trailed by a topological field theory. We exemplify these ideas in the 1+1 dimensional, chiral 3450 model which shares many of the same features. 

We comment on the effects of gauge field fluctuations on the lowest angular momentum fermion scattering states in the presence of a  magnetic monopole. While, to leading order, these zero modes can penetrate into the monopole core, in the full theory some of the zero modes are lifted and develop a small centrifugal barrier. The dynamics of the zero modes is that of a multi-flavor Schwinger model with a space-dependent gauge coupling. Symmetries and anomalies constrain the fate of the pseudo-zero modes.

}

{
\thispagestyle{empty}
}

\end{titlepage}

{\hypersetup{linkcolor=black}
\tableofcontents
}

\unitlength = .8mm

\setcounter{tocdepth}{3}

\section{Introduction and Summary}\label{sec:intro}\setcounter{footnote}{0}

The purpose of this paper is to describe some aspects of scattering of massless particles off a heavy monopole. 
Throughout, we will consider the process strictly in the realm of effective field theory. This means that our setup is a $3+1d$ $U(1)$ gauge theory with electrically charged matter, and the monopole is represented by a line operator known as the 't Hooft line. Since the underlying gauge theory has a magnetic $U(1)$ 1-form symmetry, an 't Hooft line corresponding to a charge $n$ monopole carries charge $n$ under the magnetic $1$-form symmetry and the magnetic field cannot be screened. 

Historically, this problem is motivated by the fact that grand unified theories predict the existence of stable monopoles~\cite{tHooft:1974kcl,Polyakov:1974ek}.\footnote{Furthermore, dynamical monopoles are typically present in lattice realizations of gauge theories.} Monopoles can induce proton decay at a rate that is not suppressed by the GUT scale, in disagreement with phenomenological measurements, thus ruling out the most naive versions of such models (see e.g.~\cite{doi:10.1063/1.2947547,VARubakov_1988} for early reviews). The study of scattering of charged particles off non-dynamical monopoles is a very helpful first step in understanding gauge theories with dynamical monopoles. For example the symmetries of theories with dynamical monopoles can be elucidated by first understanding the scattering off 't Hooft lines.

Most importantly for us, in the study of monopole scattering the authors of~\cite{Callan:1982ah,Rubakov:1982fp} stumbled upon a rather confusing observation. Even though the in-states are perfectly standard wave packets made out of the charged particles, the outgoing states can apparently have unconventional, and sometimes fractional,  quantum numbers. Multiple explanations have appeared in the literature for this interesting result, some of which are:

\begin{itemize} 
\item Reference \cite{PhysRevD.28.876} advocates for a probabilistic interpretation, to wit: a fractionalized excitation means that the corresponding excitation would be emitted with suppressed probability but when observed it would always have standard quantum numbers.
\item It was also suggested~\cite{PhysRevLett.52.1755,doi:10.1063/1.34591,doi:10.1146/annurev.ns.34.120184.002333} that the unconventional out-states are an artifact of the massless limit, and were we to consider a more realistic model with finite masses, the exotic states would become unstable and  decay to ordinary states in the Fock space. 
\item Another standard point of view is that the out-state should be thought of as a sort of coherent state, for which it makes no sense to talk about occupation numbers, and the confusing nature of the out-state is just a reflection of the non-trivial structure of the vacuum~\cite{Polchinski:1984uw,doi:10.1063/1.34591}. In more modern language,~\cite{Brennan:2021ewu} points out that fermion numbers are anomalous, which implies that the vacuum carries fractional charge.

\item Reference \cite{Csaki:2022qtz} questions the validity of the computation that leads to fractional quantum numbers: it is usually assumed that only the $s$-wave modes can get into contact with the monopole (while the other modes are repelled by a centrifugal potential) but perhaps some small mixing with the higher angular momentum modes can modify the conclusions. It is suggested that a quantum field theoretic treatment in terms of the soft photon dressed states of \cite{Csaki:2022tvb} would shed light on such a modification.

\end{itemize}

 Here we analyze the following simple scenario for the scattering of massless fermions off a monopole: 

\medskip

\begin{mdframed}
The outgoing radiation is attached to a topological surface that ends topologically on the 't Hooft line.
\end{mdframed}

 In other words, while the charge and energy are carried by particle-like excitations, there is a topological surface that carries no charge or energy that trails these excitations. The presence of the topological surface means that the outgoing radiation is in a twisted sector, and states in a twisted sector do not have to have the same quantum numbers as states in the ordinary Fock space.
(Similar ideas have already appeared in the literature. For example~\cite{Hamada:2022eiv} argues that the outgoing radiation lives in a different Fock space. Furthermore,~\cite{Kitano:2021pwt} proposes solitonic out-states, dubbed ``pancakes'', which have the same quantum numbers as regular particles, but are not single-particle excitations created by the fundamental fields. These new excitations of the system are again unstable for finite masses, and eventually decay to regular excitations.) 

This phenomenon is very familiar in $1+1$ dimensions, and we review it in an explicit free field theory model, called the 3450 model \cite{Bhattacharya:2006dc,Giedt:2007qg,Wang:2013yta,Wang:2018ugf,Smith:2019jnh,Smith:2020rru,Tong:2021phe,Zeng:2022grc}. See also~\cite{Maldacena:1995pq} which argued that, in their model, if the incoming fermions are in the Neveu-Schwarz sector, then the outgoing ones are in the Ramond sector. Interestingly, in the 3450 model, we observe the unusual feature that the scattering may convert a single fermion state to a composite (excited) state in the twisted Fock space, and vice versa.

For the radiation to be attached to a topological surface (and hence be in a twisted sector), the theory has to have a topological surface of co-dimension 1, i.e., a 0-form symmetry. We will see that in the 3450 model the 0-form symmetry is a standard $\mathbb{Z}_5$ invertible symmetry. Surprisingly, in the problem of scattering charges from monopoles, the radiation is often attached to non-invertible symmetries! Upon adding masses to the quarks (if this is possible) the topological surfaces typically disappear, i.e., the corresponding symmetries, invertible or not, are softly broken and the radiation comes out in the ordinary Fock space. Depending on the matter content and the monopole charge, we find that the radiation comes out attached to either invertible or non-invertible topological co-dimension 1 surfaces. The non-invertible topological surfaces we will see here are the $\mathcal D_{p/N}$ operators constructed in~\cite{Choi:2022jqy,Cordova:2022ieu}. Since non-invertible topological surfaces carry a topological field theory, the outgoing state contains the degrees of freedom of a topological theory in addition to particles with unconventional flavor charges.

To understand which twisted sectors are important for the scattering process, we have to understand the symmetries preserved by the monopole ('t Hooft line). The monopole requires boundary conditions for the bulk fields (the electrically charged particles and gauge fields) and the symmetries that are preserved by these boundary conditions need to be determined for our analysis. From the modern point of view of symmetries, one can say that only symmetries that do not participate in a 2-group~\cite{Sharpe:2015mja,Tachikawa:2017gyf,Cordova:2018cvg,Benini:2018reh} with the magnetic 1-form symmetry can be preserved by the boundary conditions of the 't Hooft line. As we will see, this holds for space-time symmetries, ordinary 0-form symmetries, and non-invertible 0-form symmetries. The symmetries that are respected by the boundary conditions of the monopoles remain as symmetries also when the monopoles become dynamical.

The fact that the out-state lives in a twisted sector is due to the special nature of the incoming and outgoing scattering modes of electric particles around a magnetic monopole. In classical physics, electric particles cannot penetrate into the monopole core. In quantum field theory, charged massless fermions in the lowest angular momentum channel can (at leading order in $e^2$) overcome the barrier due to their magnetic moment. Another way to understand these chiral modes which have no centrifugal barrier at leading order in $e^2$ is due to the index theorem (see for instance~\cite{Shnir:2005vvi,Brennan:2021ucy} for an analysis of the scattering states arising from massless fermions). The scattering of these modes is subtle as we will momentarily review.
In particular, their scattering is not consistent with unitarity unless the twisted sectors participate.

An 't Hooft line preserves the rotation symmetry around the origin, hence we can label states by their angular momentum $j\in\frac12\mathbb Z$. At \emph{tree level} (i.e., to leading order in $e^2$)  there is no centrifugal barrier for the lowest-lying mode in the partial wave expansion, $j=j_\text{min}$. For the unit charge monopole and charge 1 fermions, these are the well-known $s$-wave modes. Modes with $j>j_\text{min}$ angular momentum cannot reach the monopole core due to the standard centrifugal barrier, and hence their scattering is conceptually simpler.

The novelty of the $j_\text{min}$-modes is that each left-handed $4d$ Weyl fermion leads to chiral modes, namely, either purely incoming or purely outgoing, depending on the sign of its electric charge. This means, for instance, that a positive helicity electron must be purely incoming and a negative helicity electron must be purely outgoing. The number of such chiral modes depends on the monopole charge. If there are additional quantum numbers, one has to again classify the incoming and outgoing states according to those quantum numbers. In other words, fermions become effectively chiral two-dimensional fields, depending on time and the radial coordinate only. The scattering process is then described by a system of $2d$ fermions on the half-plane $r\ge0$, with all the details of the monopole and the UV data encoded into some boundary conditions at $r=0$. Scanning over all the possible boundary conditions, we can cover all the possible outcomes of different UV completions. The theory of $2d$ boundary conditions is well-developed and we will use anomalies to understand which symmetries may be preserved by the corresponding boundaries. Our most explicit examples are those where the boundary condition for the $2d$ system can be fully constructed. However, in some examples, even if the boundary state is not known in detail, we will use general arguments to fix the outgoing states in the scattering process.

\begin{figure}[h!]
\centering
\begin{tikzpicture}

\draw[thick] (0,0) circle (1.8cm);
\fill (0,0) circle (.1cm);
\foreach \x in {0,...,7} \draw[very thick,->,>=stealth] ({2.1*cos(45*\x)},{2.1*sin(45*\x)}) -- ({1.4*cos(45*\x)},{1.4*sin(45*\x)});

\begin{scope}[shift={(8,0)}]
\draw[thick] (0,0) circle (1.8cm);
\fill (0,0) circle (.1cm);
\foreach \x in {0,...,7} \draw[very thick,<-,>=stealth] ({2.1*cos(45*\x)},{2.1*sin(45*\x)}) -- ({1.4*cos(45*\x)},{1.4*sin(45*\x)});

\begin{scope}
\clip (0,0) circle (1.8cm);
\foreach \x in {-7,...,7} \draw (.3*\x-2,-3) -- (.3*\x+2,3);
\end{scope}
\fill[rounded corners,fill=white] (-.8, .13) rectangle (.8,.5) {};
\node[scale=.8,text width=2cm] at (-1,-2.3) {Outgoing radiation};
\node[scale=.8] at (0,.3) {Monopole};

\fill[rounded corners,fill=white] (-.6, -1-.1) rectangle (.6,-.55-.1) {};
\node[scale=.8] at (0,-.8-.1) {TQFT};

\end{scope}

\node[scale=.8] at (0,.3) {Monopole};
\node[scale=.8,text width=2cm] at (-1,-2.3) {Incoming radiation};

\end{tikzpicture}
\caption{On the left, a spherically symmetric wave of radiation propagates towards the monopole, at the center of the sphere. On the right, the spherically symmetric outgoing radiation is accompanied by a non-trivial $3d$ TQFT in the interior of the spherical shell.}
\label{fig:TQFT_radiation}
\end{figure}
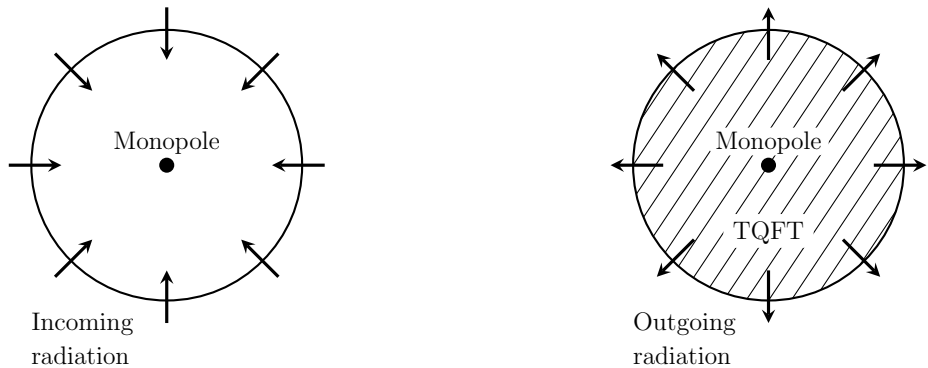

The simplest example is QED$_4$, namely a $4d$ $U(1)$ gauge theory with $N_f$ Dirac fermions of unit charge, in the presence of a monopole of charge $n\in\mathbb Z$. The dynamics of the lowest-lying partial wave modes reduces to $nN_f$ left-moving $2d$ fermions, and as many right-moving ones. There are several options for which symmetries the monopole preserves. (That is, there are several subgroups of the full 0-form symmetry group that do not participate in a 2-group with the magnetic 1-form symmetry.) For each case, we will show that if we send a left-moving fermion towards the boundary, what comes out is a right-moving one, attached to an invertible topological line that implements a discrete chiral symmetry on the fermions. For one natural choice of subgroup, that often arises from common UV completions and which exhibits the main features we want to elucidate in this note, the topological line acts as $\psi_R\mapsto e^{-4\pi i/nN_f}\psi_R$. This line generates an invertible symmetry in the effective $2d$ system. Lifting this result to $4d$, the conclusion is that an incoming $j_\text{min}$-wave becomes an outgoing one, attached to a topological operator that acts as $\Psi\mapsto e^{2\pi i \gamma^\star/nN_f}\Psi$ on the fermions. While there is no invertible symmetry in $4d$ that acts this way (for $n>2$), there is in fact a non-invertible one that does, namely the operator $\mathcal D_{2/n}$.\footnote{In the notation of~\cite{Choi:2022jqy}, our operator has $p/N=2/n$. In the notation of~\cite{Cordova:2022ieu} we use $k=nN_f/2$.} The conclusion is that the out-state lives in a Hilbert space twisted by this (generally non-invertible) defect, see figure~\ref{fig:TQFT_radiation}. We will see that the outgoing states that we propose preserve the necessary symmetries. For the discussion above, the following fact is important: if we restrict our attention to the $\mathbb{Z}^{(1)}_n$ subgroup of the $U(1)^{(1)}$ 1-form symmetry, then the  $\mathcal D_{p/n}$ surfaces do not participate in a 2-group. In practice, this also means that if we were to make charge $n$ monopoles dynamical, then the symmetries $\mathcal D_{p/n}$ would survive. We will see this explicitly when we study the boundary conditions for the fermions near the monopole.

A general question is how to directly detect the existence of a topological surface that trails the outgoing radiation experimentally. This seems difficult as it requires measuring the non-trivial monodromy induced by the topological surface. This can be done, for example, with an out-of-time ordered correlation function as in~\cite{Chang:2020imq}.

Next, we discuss some basic properties of the corrections to the above tree-level picture. The $2d$ massless fermions arise due to the index theorem in the background of a magnetic monopole. Some of the zero modes can be lifted due to gauge field fluctuations. That means that a small centrifugal barrier can develop at subleading orders in the expansion in $e^2$ (more precisely, some modes develop a barrier in perturbation theory in $e^2$ and some modes are only subject to non-perturbative corrections). For the minimal case of a single Dirac fermion propagating around the minimal 't Hooft line, a small centrifugal barrier of order $e^2$ develops due a variant of the Schwinger model on the $(r,t)$ half-plane. Additionally, in the vacuum, the fermion bilinear condenses with a condensate that decays as $1/r$. For two units of the minimal monopole flux, one finds one mode that develops a centrifugal barrier at order $e^2$ and another mode at the Berezinskii-Kosterlitz-Thouless (BKT) point~\cite{Berezinsky:1970fr,J_M_Kosterlitz_1973} whose fate needs to be further analyzed (it may develop an exponentially small symmetry-breaking condensate and barrier, depending on a sign we have not determined\footnote{We thank T.~Dumitrescu for a discussion about this.}). For the multi-flavor case, some modes with an exactly vanishing centrifugal barrier must remain. We discuss a simple scenario that matches the anomalies (though at present the evidence for this scenario is essentially non-existent), which is a broken $\mathbb{Z}_n$ chiral symmetry due to a condensate around the monopole and an $SU(N_f)_n$ WZW model in each of the vacua.

The outline of the paper is as follows. In section~\ref{sec:2dBCFT} we review the physics of a very elementary $2d$ model, the so-called 3450 system. It contains a lot of the conceptual points. In section~\ref{sec:non-chiral_4d} we discuss the lowest-lying wave dynamics of charged particles propagating in the background of a magnetic monopole. We first perform the analysis to leading order in $e^2$ and then discuss how pseudo-zero-modes develop a small centrifugal barrier as various corrections are included. In section~\ref{sec:uplift_4d} we uplift the $2d$ topological lines to generalized symmetry surfaces in $4d$ and discuss the consequences of having a topological theory on the surface attached to the outgoing radiation. In section~\ref{sec:chiral} we discuss chiral theories and list a few open questions. Finally, we record some technical details in a few appendices.

\section{$\boldsymbol{2d}$ BCFT}
\label{sec:2dBCFT}

As outlined in the introduction, scattering processes in $4d$ that involve heavy monopoles reduce, at low energies and at leading order in the fine structure constant, to an effective $2d$ problem with massless non-interacting fermions. 
We will study such theories in this section. The two-dimensional fermions propagate in the half plane $t\in\mathbb R$, $r\ge0$ and they are free in the bulk, while the entirety of the $4d$ information about the interactions is encoded in some boundary conditions at $r=0$, where the monopole sits. For a given UV completion of the $4d$ theory, the boundary conditions can in principle be determined. Instead, we would like to characterize the space of consistent boundary conditions from the infrared effective theory point of view. The symmetries and anomalies will prove enough for our goals in several examples. Let us, then, begin by reviewing some relevant facts about boundaries, symmetries, and anomalies in QFT.

It is an open problem to determine when a quantum field theory can be placed on a manifold with a boundary. We may further require that the boundary condition is compatible with energy conservation and we may require additional internal and space-time symmetries to be preserved by the boundary. An important statement~\cite{Jensen:2017eof,Thorngren:2020yht} is that for a theory with internal symmetry $G$ with an 't Hooft anomaly there exists no simple (we often omit the adjective ``simple'' in the rest of the paper) boundary condition preserving $G$. However, depending on the 't Hooft anomaly, a subgroup of $G$ may be preserved.\footnote{We have to explain what it means for a boundary condition to preserve a symmetry. For our purposes, it will be sufficient to require that the symmetry co-dimension 1 topological surface has a topological endpoint on the boundary. This is sufficient, for instance, to declare that there is a conserved charge in the presence of the boundary. Alternative definitions are discussed in~\cite{Choi:2023xjw}.}

In two dimensions, for continuous symmetries, it is easy to prove that symmetry-preserving conformal boundaries exist only for symmetries with no 't Hooft anomalies. As argued by Cardy~\cite{Cardy:1989ir}, a boundary $\mathcal B$ preserves some chiral algebra if and only if the currents reflect smoothly off the boundary (i.e., no charge is deposited on the boundary), namely $(J_n+\tilde J_{-n})|\mathcal B\rangle=0$. Such a constraint is inconsistent unless the operators $J_n+\tilde J_{-n}$ all commute with each other. But the commutators of such currents are given by the standard Schwinger term $n\delta_{n+n',0}$, whose coefficient is nothing but the 't Hooft anomaly. Hence, only anomaly-free symmetries can possibly be preserved by $\mathcal B$.
For example, for the Virasoro algebra the symmetry condition is $(L_n-\tilde L_{-n})|\cB\rangle=0$. It is clear that all the operators $L_n-\tilde L_{-n}$ commute with each other if and only if $c_L=c_R$. This is just the familiar statement that, if the theory has a gravitational anomaly, the boundary cannot preserve conformal invariance. Note an important fact that follows from considering the $n=0$ components of the symmetry conditions: the boundary cannot store charge or energy. This will be important below, because it implies that all the charge and energy must come out at infinity.

In general it is extremely hard to find symmetric states that also solve the Cardy conditions~\cite{Cardy:1989ir}, necessary for a consistent open-channel spectrum, and explicit answers are mostly known for theories that are rational with respect to the symmetry preserved by the boundary  (for a review see~\cite{Cardy:2004hm}). The reason is that for these theories, and for these only, the number of Ishibashi states is finite.

Let us now move on to the case that is of interest to us, namely $2d$ free fermions. To be concrete, let us consider the theory of $N$ free, massless left- and right-moving complex fermions $\psi_i(z)$, $\tilde \psi_i(\bar z)$ with $i=1,\dots, N$, whose action is given by 
\be 
S=\frac{1}{2\pi} \int d^2z \lb \psi_i \partial \psi^\dagger_i+\tilde \psi_i \bar \partial \tilde{\psi}^\dagger_i \rb\,.
\ee
Away from the boundary, the fermions have a chiral symmetry
\be
SO(2N)_L \times SO(2N)_R\,,
\ee
(ignoring discrete factors). If we introduce a boundary $\cB$ in this system, we must impose boundary conditions that relate the left- and right-moving fermions. We may require this boundary to preserve, for example, a diagonal $SO(2N)$ algebra. Given that the theory is holomorphic with respect to this chiral algebra, there is essentially a unique Cardy state for this symmetry.\footnote{In terms of the Majorana components of $\psi,\tilde\psi$, the boundary condition becomes $\chi_i=M_{ij}\tilde\chi_j$ for some orthogonal matrix $M$ that specifies the embedding $SO(2N)\subset SO(2N)_L\times SO(2N)_R$.}

Rather than preserving the full diagonal $SO(2N)$ symmetry, we may require the conservation of a smaller chiral algebra, e.g.~some subgroup $U(1)^r\subset SO(2N)_L\times SO(2N)_R$. For generic $r$ it is not possible to find the explicit Cardy state but, for $r=N$, the theory is rational with respect to $U(1)^N$ (although no longer holomorphic), and therefore the boundary state can again be written explicitly. More generally, Cardy states can only be written explicitly for boundary conditions that preserve a chiral algebra that embeds into $SO(2N)_1$ conformally. 
The boundary states preserving a general embedding of $U(1)^N$ were constructed in \cite{Smith:2019jnh}.

An embedding of $U(1)^r_L\times U(1)^r_R\subset SO(2N)_L\times SO(2N)_R$ can be specified by defining the currents that generate the $U(1)$ subgroups. Concretely, we can choose integral matrices $Q,\tilde Q$ and define currents
\begin{equation}
\mathcal J_a=Q_{ai}\psi_i\psi^\dagger_i,\qquad \tilde{\mathcal J}_a=\tilde Q_{ai}\tilde \psi_i\tilde{\psi}^\dagger_i,\qquad a=1,\dots,r\,.
\end{equation}
These currents generate $U(1)^r$ chiral algebras with $K$-matrices $Q^tQ$ and $\tilde Q^t\tilde Q$. We can require that the boundary preserves a diagonal $U(1)^r$ chiral algebra by setting
\begin{equation}\label{eq:bc_current}
(\mathcal J_{a,n}+\tilde{\mathcal J}_{a,-n})|\mathcal B\rangle=0\,.
\end{equation}
The existence of a boundary state $|\cB\rangle$ is predicated on the vanishing of the commutator $[\cJ_{a,n}+\tilde \cJ_{a,-n},\cJ_{a',n'}+\tilde \cJ_{a',-n'}]$ which holds if and only if
\be 
Q^tQ=\tilde Q^t\tilde Q\,.
\ee
This is precisely the condition that there are no mixed 't Hooft anomalies within the diagonal $U(1)^r$.

Let us discuss these boundary conditions from the point of view of the bosonized theory.
Consider a free compact boson $X(z, \bar z) \sim X(z, \bar z)+2\pi$ at the radius which is dual to a complex fermion. The action is\footnote{Our convention is $d^2 z= 2 d\sigma^1 d \sigma^2$ and $\partial_z=\frac{1}{2}(\partial_1 -i \partial_2)$.}
\be 
\label{eq:Sfreeboson}
S=\frac{1}{4\pi} \int d^2z\, \partial X \bar \partial X\,.
\ee 
 The equation of motion for $X$ is solved by introducing chiral bosons $X(z,\bar z)=X(z)+\tilde X(\bar z)$.
The well-defined vertex operators are of the form $e^{i(k+w/2)X(z)}e^{i(k-w/2)\tilde X(\bar z)}$ with $k,w \in \Z$. Chiral vertex operators are obtained for example by choosing $k=1$, $w=\pm 2$ corresponding to $e^{2i X(z)}$, and $e^{2i \tilde X(\bar z)}$. The scaling dimension and spin of the general exponential vertex operator are given by
\be 
\Delta =k^2+\frac{1}{4}w^2 ~,\quad S= kw\,. 
\ee
The bosonic theory is similar but not identical to the original fermion theory. First of all, the vertex operator which is identified with the original fermions is  
$\psi(z) \cong e^{i X(z)}$
and likewise for the right-mover. These operators do not correspond to  $k,w \in \Z$ and should be interpreted as being attached to a topological line in the bosonic theory. Conversely, the bosonic theory has well-defined vertex operators such as the non-chiral $k=0, w=1$ which in the fermionic theory is interpreted as being attached to a line.

The boundary condition~\eqref{eq:bc_current}, for a boundary along the $\sigma^1$ direction, can be written as
\begin{equation}
(Q\partial X-\tilde Q\bar\partial\tilde X)\biggr|_{\mathcal B}=0~.
\end{equation}
In general there could be many solutions and the classification is not known.
For $r=N$ the matrices $Q,\tilde Q$ are $N\times N$, the theory is rational and we can explicitly find a boundary state by imposing a more restrictive condition
\begin{equation}\label{eq:bc_bosonized}
(QX-\tilde Q\tilde X)\biggr|_{\mathcal B}=0~.\end{equation}
(There could be nonzero constants on the right hand side of~\eqref{eq:bc_bosonized}, which are exactly marginal boundary parameters. We set them to zero for simplicity.)
Since $Q$ and $\tilde Q$ are full rank we can invert~\eqref{eq:bc_bosonized} and write $X=\mathcal R\tilde X$ with $\mathcal{R}:=Q^{-1}\tilde Q$ an orthogonal matrix. We can then follow the evolution of excitations by simply matching up the values of $X$ and $\tilde X$ at the boundary. (See, for example \cite{Yegulalp:1994eq} for an early application of this to the Callan-Rubakov problem.)

Since $\mathcal{R}$ is, in general, a matrix over the rationals, excitations created by acting on the vacuum with well-defined vertex operators can evolve to fractional excitations. Therefore, the scattering problem inevitably mixes up local states and states attached to topological lines. The example below demonstrates these ideas.

\subsection{The 3450 Model}\label{sec:345}

In this section we study a $2d$ toy model that nicely encapsulates some of the non-trivial physics of $4d$ monopole scattering.
Consider a two-dimensional system consisting of two left-moving and two right-moving free complex fermions $\psi_1,\psi_2,\tilde\psi_1,\tilde\psi_2$ on the half plane $(r,t)\in \mathbb R_+\times\mathbb R$,
\begin{equation}
\begin{tikzpicture}[baseline=.9cm]

\draw[very thick] (-3,0) -- (-3,2);
\draw[->,>=stealth,thick] (-1.2, 0.2) -- (-2.2, 0.7);
\draw[<-,>=stealth,thick] (-1.2, 1.8) -- (-2.2, 1.3);
 
\foreach \x in {0,...,12} \draw (-3-.2,.15*\x+.04) -- (.2-3-.2,.15*\x+.1+.04);
\foreach \x in {-1,...,11} \draw (-3-.1,.15*\x+.09+.075) -- (.2-3-.2,.15*\x+.1+.04+.075);
 
\node at (-0.9, 1.8) {$\tilde\psi_i$};
\node at (-1.7+.8,.1) {$\psi_i$};
 
\filldraw (-1.2, 0.2) circle (1pt);
\filldraw (-2.2, 1.3) circle (1pt);
 
\end{tikzpicture}
\end{equation}
 
At the boundary $r=0$ we impose some conformal boundary condition, the details of which we shall discuss below. 

The bulk flavor symmetry acting on the fermions is $SO(4)_L\times SO(4)_R$ although this group has an 't Hooft anomaly and therefore cannot be preserved by the boundary. In this toy model we will choose to preserve (at least) a $U(1)\subset SO(4)_L\times SO(4)_R$ subgroup where the left-movers carry charges $3$ and $4$, while the right-movers carry charges $5$ and $0$. This subgroup is anomaly-free since $3^2+4^2=5^2+0^2$, and therefore it can be preserved by the boundary in a conformal invariant way. While $\tilde\psi_2$ is neutral under $U(1)$, the main reason we include it here is so that the gravitational anomaly vanishes, $c_L-c_R=2-2\equiv0$ (otherwise one cannot hope to find a conformal boundary condition).
 
Let us declare that the boundary preserves this $U(1)$ symmetry. What happens if we throw the left-moving fermion with charge $3$ onto the boundary? The most natural answer is that the out-state will be some superposition of the right-movers. But here we encounter a puzzle. Since the right-moving fermions have charges $5$ and $0$, there exists no state in their Fock space with charge $3$. So it seems impossible to write an out-state consistent with the $U(1)$ symmetry. More generally, if we throw in $n_1$ charge $3$ fermions and $n_2$ charge 4 fermions then we would need
\begin{equation}\label{eq:wrong_345_out}
\tilde n_1=\tfrac35n_1+\tfrac45n_2
\end{equation}
outgoing charge 5 fermions, and since this expression is not in general integral the statement is meaningless. Similar fractional combinations of particles appeared in the discussion of scattering off monopoles in $4d$, which led to the various interpretations cited in the introduction. 

In the following, we will discuss the resolution of this puzzle. We will begin by studying an example of an explicit boundary state where the answer can be presented in full detail and then comment on more general cases. 

\subsection{An Explicit Boundary State}\label{sec:u1xu1}

The outcome of any scattering process off the boundary is determined by the choice of boundary state. The $U(1)$ symmetry above does not embed conformally into $SO(4)_1$, which usually means that it is very hard to classify all solutions to the Cardy conditions. To make some progress, it is useful to impose the conservation of a second $U(1)$ symmetry. Now, $U(1)\times U(1)'$ does embed conformally into $SO(4)_1$ (i.e., the free fermions are rational with respect to $U(1)\times U(1)'$), and therefore the Cardy states can be easily constructed.

Up to a change of basis, there is a unique choice for the second symmetry that makes $U(1)\times U(1)'$ free of 't Hooft anomalies, and it takes the form
\begin{equation}
\begin{array}{c|cccc}
&\psi_1&\psi_2&\tilde\psi_1&\tilde\psi_2\\ \hline
U(1)&3&4&5&0\\
U(1)'&4&-3&0&5
\end{array}
\end{equation}
In other words, we choose charge matrices
\begin{equation}
Q=\begin{pmatrix}3&4\\4&-3\end{pmatrix},\qquad\tilde Q=\begin{pmatrix}5&0\\0&5\end{pmatrix}\,,
\end{equation}
which indeed satisfy $Q^tQ=\tilde Q^t\tilde Q$. We impose the boundary condition~\eqref{eq:bc_bosonized}, to wit,
\begin{equation}
X\bigr|_{\mathcal B}=\mathcal R\tilde X\bigr|_{\mathcal B},\qquad \mathcal R=Q^{-1}\tilde Q\equiv\frac15\begin{pmatrix}3&4\\4&-3\end{pmatrix}\,,
\end{equation}
which completely determines the scattering process. If we bring the left-moving field $e^{iX_i}$ towards the boundary, what comes out is the right-moving field $e^{i\mathcal R_{ij}\tilde X_j}$; for example, if we throw the left-mover $e^{iX_1}$, it scatters off the boundary and becomes
\begin{equation}\label{eq:345_X1_X2}
\begin{aligned}
e^{iX_1}&\longrightarrow e^{i\frac35\tilde X_1+i\frac45\tilde X_2}\,.
\end{aligned}
\end{equation}
Note that the out-states are fractional vertex operators, i.e., they are not invariant under $\tilde X_i\to\tilde X_i+2\pi$. This means that their correlation functions have branch cuts, that is, they live at the end of a topological line. Therefore, the resolution of the ``paradox'' about the outgoing states is rather clear: the scattering turns local operators into twist operators.\footnote{This is in fact a natural answer for the following reason. If we unfold the theory (cf.~\eqref{eq:unfold_345} below), the boundary becomes a topological interface. It is a generic feature for local operators to become twist operators as we drag them across a topological interface, see e.g.~\cite{Frohlich:2004ef,Chang:2018iay,Choi:2021kmx} for some examples.} Equivalently, the $S$-matrix maps the regular Fock space into a twisted Hilbert space:
\begin{equation}
\begin{tikzpicture}[baseline=.9cm]
 
\draw[very thick] (-3,0) -- (-3,2);
\draw[->,>=stealth,thick] (-1.2, 0.2) -- (-2.2, 0.7);
\draw[<-,>=stealth,thick] (-1.2, 1.8) -- (-2.2, 1.3);
\draw[decorate, decoration={snake,amplitude=.6mm}] (-3,1) -- (-2.2, 1.3);
 
\foreach \x in {0,...,12} \draw (-3-.2,.15*\x+.04) -- (.2-3-.2,.15*\x+.1+.04);
\foreach \x in {-1,...,11} \draw (-3-.1,.15*\x+.09+.075) -- (.2-3-.2,.15*\x+.1+.04+.075);
 
\node at (-.1, 1.9) {$e^{i\frac35\tilde X_1+i\frac45\tilde X_2}$};
\node at (-1.7+1,.1) {$e^{iX_1}$};
 
\filldraw (-1.2, 0.2) circle (1pt);
\filldraw (-2.2, 1.3) circle (1pt);

\end{tikzpicture}
\end{equation}

The branch cut defines a symmetry defect in our system. The action of this symmetry on the basic fields can be determined simply by computing the monodromy of the out-state $e^{i\frac35\tilde X_1+i\frac45\tilde X_2}$ around the different local operators of the theory. It is clear that this twist operator is local to $e^{iX_1}$ and $e^{iX_2}$, and it has monodromy $3/5$ and $4/5$ around $e^{i\tilde X_1}$ and $e^{i\tilde X_2}$, respectively. This means that the topological line implements a $\mathbb Z_5$ symmetry under which the fields $e^{iX_1},e^{iX_2},e^{i\tilde X_1},e^{i\tilde X_2}$ carry charges $0,0,3,4$, respectively. The general claim is, then, that the out-state of the scattering process lives in a $\mathbb Z_5$-twisted sector, i.e., it is attached to the topological line that implements this symmetry.
This $\mathbb Z_5$ symmetry is embedded inside $U(1)\times U(1)'$, which are the symmetries that are preserved by the boundary and can end topologically on it.

More generally, if the in-state contains $n_1$ copies of $X_1$ and $n_2$ copies of $X_2$, then the out-state is
\begin{equation}
e^{in_1X_1+i n_2X_2}\longrightarrow e^{i(\frac35n_1+\frac45n_2)\tilde X_1+i(\frac45n_1-\frac35n_2)\tilde X_2}\,.
\end{equation}
Next, we will interpret this expression in the fermionic language. In order to make the translation as seamless as possible, it is convenient to rewrite the out-state in a slightly different way, namely we want to write this operator as a \emph{local} excitation sitting on top of the \emph{vacuum state} of the $\mathbb Z_5$-twisted Hilbert space. In practice, this means that we will ``add and subtract'' a local operator
\begin{equation}
e^{i(\frac35n_1+\frac45n_2)\tilde X_1+i(\frac45n_1-\frac35n_2)\tilde X_2}=e^{i\tilde n_1\tilde X_1+i\tilde n_2\tilde X_2}\times e^{i(\frac35n_1+\frac45n_2-\tilde n_1)\tilde X_1+i(\frac45n_1-\frac35n_2-\tilde n_2)\tilde X_2}\,,
\end{equation}
where $\tilde n_1,\tilde n_2$ are two integers. The first factor, $e^{i\tilde n_1\tilde X_1+i\tilde n_2\tilde X_2}$, is properly quantized and hence it describes a local operator. We will interpret the second factor as the vacuum state of the twisted sector. The integers $\tilde n_1,\tilde n_2$ are defined by the condition that the scaling dimension of the vacuum be as small as possible, namely we want to minimize
\begin{equation}
\min_{\tilde n_1,\tilde n_2\in\mathbb Z}\tfrac12\bigl(\tfrac35n_1+\tfrac45n_2-\tilde n_1\bigr)^2+\tfrac12\bigl(\tfrac45n_1-\tfrac35n_2-\tilde n_2\bigr)^2\,,
\end{equation}
with solution
\begin{equation}\label{eq:out_tilden_1}
\begin{aligned}
\tilde n_1&=\bigl\lfloor\tfrac35n_1+\tfrac45n_2+\tfrac12\bigr\rfloor\,,\\
\tilde n_2&=\bigl\lfloor\tfrac45n_1-\tfrac35n_2+\tfrac12\bigr\rfloor\,.
\end{aligned}
\end{equation}
 For instance, the process~\eqref{eq:345_X1_X2} can also be written as
 \begin{equation}\label{eq:345_X1_X2_local}e^{iX_1}\longrightarrow e^{i\tilde X_1+i\tilde X_2}\times\text{twist-vacuum}\,,
 \end{equation}
 where ``twist-vacuum'' denotes the operator $e^{-i\frac25\tilde X_1-i\frac15\tilde X_2}$, which is the vertex operator in the $\mathbb Z_5$-twisted sector with the lowest scaling dimension.

We now switch back to the fermionic language. We learned from the bosonized picture that the out-state is typically a twist operator attached to a topological line that generates a $\mathbb Z_5$ symmetry under which the fermions carry charges 
\begin{equation}
\begin{array}{c|cccc}
&\psi_1&\psi_2&\tilde\psi_1&\tilde\psi_2\\ \hline
\Z_5&0&0&3&4
\end{array}
\end{equation}
Let $T$ denote the generator of this discrete symmetry, and $|T^k\rangle$ denote the vacuum state of the Hilbert space twisted by $T^k$. Then an arbitrary state in this Hilbert space can be written as $\psi_1^{n_1} \psi_2^{n_2}\cdots |T^k\rangle$ for some integers $n_i$. We write $T^k+n_1 \psi_1+n_2 \psi_2 +\cdots$ as a shorthand for the operator that maps to this state under the state-operator correspondence. Then the fermionic analogue of~\eqref{eq:345_X1_X2_local} is
\begin{equation}\label{eq:basic_345_fermion}
\psi_1\longrightarrow T+\tilde\psi_1+\tilde\psi_2\,.
\end{equation}
The operator $\psi_1$ carries charge $(3,4)$ under $U(1)\times U(1)'$, while $\tilde\psi_1$ carries charge $(5,0)$ and $\tilde\psi_2$ charge $(0,5)$; hence, the claim will be that the endpoint of $T$ carries $(-2,-1)$ units of charge so that~\eqref{eq:basic_345_fermion} conserves $U(1)\times U(1)'$.

\begin{figure}[h!]
\centering
\begin{tikzpicture}
 
\filldraw (0,0) circle (2pt);
\draw[thick,decorate, decoration={snake,amplitude=.6mm}] (0,0) -- (2,0);
 
\foreach \x in {1,...,4} \draw[very thin, dashed] (0,0) circle (.4*\x cm);
 
\node at (3,0) {$\cong$};
 
\draw (4.5,-1.3) -- (4.5,1.3);
\draw (6,-1.3) -- (6,1.3);
\draw (5.25,1.3) ellipse (0.75cm and .3cm);
\draw (5.25-.75,-1.3) arc(180:360:.75cm and .3cm);
 
\foreach \x in {1,...,5} \draw[very thin, dashed] (5.25-.75,-1.3+.43*\x) arc(180:360:.75cm and .3cm);
 
\draw[thick,decorate, decoration={snake,amplitude=.6mm}] (5.25,-1.6) -- (5.25,1);
 
\end{tikzpicture}
\caption{Radial quantization on the plane is equivalent to a cylinder geometry by a conformal transformation. The topological line implements a twist in the Hilbert space of the fermions living at its endpoint.}
\label{fig:twist_H}
\end{figure}
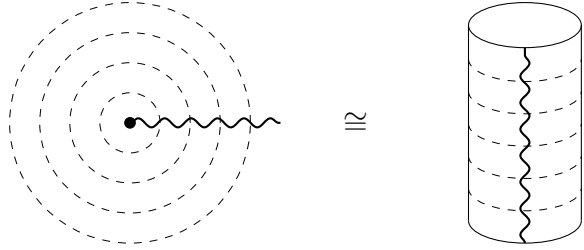

Let us compute the $U(1)\times U(1)'$ charge carried by the endpoint of the $\mathbb Z_5$ topological line. For future reference, we consider a more general setup where we have a bunch of fermions with charges $Q_i$ under a flavor $U(1)$ symmetry, and charges $q_i$ under a flavor $\mathbb Z_n$ symmetry. Let the elements of $\mathbb Z_n$ be $T^k$ with $k=0,1,\dots,n-1$, such that $T^n=1$. By the state-operator correspondence, the spectrum of operators that live at the end of $T^k$ is isomorphic to the $S^1$ Hilbert space defined by the boundary conditions\footnote{The minus sign is because we look at Neveu-Schwarz but of course the Ramond sector is qualitatively identical, up to stacking the $T^k$ line with $(-1)^F$.} (see figure~\ref{fig:twist_H})
\begin{equation}
\label{eq:twist}
\psi_i(\sigma+2\pi)=-e^{2\pi i q_i k/n}\psi_i(\sigma)\,.
\end{equation}
 
Next, we ask --- what is the $U(1)$ charge of the states in this Hilbert space? The creation operators have charges $Q_i$ so of course the non-trivial question is what is the charge of the vacuum state $|T^k\rangle$. This state is a Dirac sea of modes with positive momentum, so each fermion with boundary condition $\psi(\sigma+2\pi)=e^{2\pi i\eta}\psi(\sigma)$ contributes a vacuum charge\footnote{We assume that $\eta\neq0$ for otherwise there are zero-modes that need to be treated separately.}
\begin{equation}\label{vacuumcharge}
\sum_{p\in\mathbb N+[\eta]}1=\sum_{p\in\mathbb N+[\eta]}p^{-s}\biggr|_{s\to0}=\zeta(s,[\eta])\biggr|_{s\to0}=\frac12-[\eta]\,,
\end{equation}
where $[\eta]:=\eta-\lfloor\eta\rfloor$ denotes the fractional part of $\eta$, i.e., such that $0<[\eta]<1$. Finally, using $\eta_i=\frac12+q_ik/n$ and summing over all fermions, we get a total vacuum charge\footnote{As a quick check, note that the fractional part of $Q_\text{vac}$ is $-\frac{k}{n}\sum_i Q_iq_i\gamma_i^\star$, namely it is given by the mixed anomaly between $U(1)$ and $\mathbb Z_n$. If there is no mixed anomaly, then $Q_\text{vac}$ is an integer. This is as expected, since if $Q_\text{vac}$ is fractional, then the $U(1)$ symmetry is realized projectively on the $\mathbb Z_n$-twisted Hilbert space.}
\begin{equation}\label{eq:general_Qvac}
Q_\text{vac}=\sum_i Q_i\times\bigl(\tfrac12-\bigl[\tfrac12+\tfrac1nq_ik\bigr]\bigr)\gamma_i^\star\,,
\end{equation}
where $\gamma_i^\star=\pm1$ for left- and right-movers, respectively. This is the $U(1)$ charge carried by the endpoint of the $\mathbb Z_n$ topological line. 
In other words, the charge of $\psi_1^{n_1}\psi_2^{n_2}\cdots |T^k\rangle$ is $Q_\text{vac}+n_1Q_1+n_2Q_2+\cdots$.

We can now apply this to the 3450 model. Using $Q_i=3,4,5,0$ under the first $U(1)$, and $Q'_i=4,-3,0,5$ under the second $U(1)$, and $q_i=0,0,3,4$ under the $\mathbb Z_5$ twist symmetry, we find that $T^k$ adds $(Q_\text{vac},Q'_\text{vac})$ units of $U(1)\times U(1)'$ charge to the operators that live at the end of the line, where
\begin{equation}
\begin{aligned}
Q_\text{vac}&=-5\bigl(\tfrac12-\bigl[\tfrac12+\tfrac35k\bigr]\bigr)=\{0,-2,1,-1,2\}\\
Q'_\text{vac}&=-5\bigl(\tfrac12-\bigl[\tfrac12+\tfrac45k\bigr]\bigr)=\{0,-1,-2,2,1\}
\end{aligned}
\end{equation}
for $k=\{0,1,2,3,4\}$, respectively. In particular, $T$ adds $(-2,-1)$ units of charge, and~\eqref{eq:basic_345_fermion} indeed conserves $U(1)\times U(1)'$.

More generally, if the in-state consists of $n_1$ copies of $\psi_1$ and $n_2$ copies of $\psi_2$, the out-state is
\begin{equation}\label{eq:out_state_345}
n_1\psi_1+n_2\psi_2\longrightarrow T^{n_1+3n_2}+\tilde n_1\tilde\psi_1+\tilde n_2\tilde\psi_2\,,
\end{equation}
where $\tilde n_1,\tilde n_2$ are given by~\eqref{eq:out_tilden_1}. 
Using~\eqref{eq:general_Qvac} it is easily checked that~\eqref{eq:out_state_345} conserves $U(1)\times U(1)'$. The interpretation of this result is, as before, that the out-state consists of $\tilde n_1$ copies of the local operator $\tilde\psi_1$ and $\tilde n_2$ copies of the local operator $\tilde\psi_2$, sitting on top of a $\mathbb Z_5$-twisted vacuum state, which lives at the end of a topological line.

Let us now briefly come back to the question of the fractionalized quantum numbers of the outgoing states. We see that $U(1)\times U(1)'$ is preserved and all the states have integer quantum numbers under it. However, the fermion numbers are fractional. Indeed, suppose we tried to count the number of $\tilde\psi_1$ or $\tilde\psi_2$ modes individually, i.e., the charges associated to the currents $\tilde\psi_i^\dagger\tilde\psi_i$ (no sum over $i$). 
In~\eqref{eq:out_state_345}, the term $\tilde n_1\tilde\psi_1$ contributes $\tilde n_1$ units to the charge $\tilde\psi_1^\dagger\tilde\psi_1$, but there is also a contribution from the line $T^k$. Using our general result~\eqref{eq:general_Qvac}, we learn that $T^k$ adds $(-\frac12+[\frac12+\frac35k])$ units of $\tilde\psi_1^\dagger\tilde\psi_1$-charge. All in all, the occupation number of $\tilde\psi_1$ on the out-state is
\begin{equation}
\tilde n_1+(-\tfrac12+[\tfrac12+\tfrac35(n_1+3n_2)])=\tfrac35n_1+\tfrac45n_2\,,
\end{equation}
which agrees with the naive answer~\eqref{eq:wrong_345_out}. In conclusion, the charge of $\tilde\psi_i^\dagger\tilde\psi_i$ on the out-state is typically fractional. The naive explanation is that the out-state consists of a fractional number of $\tilde\psi_i$ operators while the more precise explanation is that the out-state consists of an integral number of $\tilde\psi_i$ operators, attached to a topological line. This topological line has a mixed anomaly with the current $\tilde\psi_i^\dagger\tilde\psi_i$, so it adds a fractional part to the charge of the out-state. Due to the mixed 't Hooft anomaly, $\tilde\psi_i^\dagger\tilde\psi_i$ is not conserved in the scattering process (this is the 1+1 dimensional analog of the non-conservation of particle number, such as baryon number, in scattering off a monopole). By contrast, $U(1)\times U(1)'$ has no mixed anomalies and hence does not fractionalize and is conserved.

\subsection{Less Symmetric Boundary States}\label{sec:u1z5andbeyond}

Above we have discussed conformal boundary states that preserve $U(1)\times U(1)'$. However, the puzzle we are considering arises for less symmetric boundary states as well, so we must examine these. In particular, we would like to examine putative conformal boundary states which do not preserve the crucial $\Z_5$ symmetry from the previous subsection.

Let us begin by discussing a further constraint on the out-state on top of $U(1)$ invariance, namely the fact that the boundary is conformal. To this end, we note that our theory can be written as $\text{CFT}_1\times\overline{\text{CFT}_2}$, where $\text{CFT}_1$ denotes the left-moving free fermions $\psi_i$, and $\overline{\text{CFT}_2}$ the right-moving ones $\tilde\psi_i$. This means that the theory can be unfolded into the plane $\mathbb R^2$, with all fermions moving to the left:
\begin{equation}\label{eq:unfold_345}
\begin{tikzpicture}[baseline=.9cm]

\draw[very thick] (-3,0) -- (-3,2);
\draw[->,>=stealth,thick] (-1.2, 0.2) -- (-2.2, 0.7);
\draw[<-,>=stealth,thick] (-1.2, 1.8) -- (-2.2, 1.3);
 
\foreach \x in {0,...,12} \draw (-3-.2,.15*\x+.04) -- (.2-3-.2,.15*\x+.1+.04);
\foreach \x in {-1,...,11} \draw (-3-.1,.15*\x+.09+.075) -- (.2-3-.2,.15*\x+.1+.04+.075);
 
\node at (-0.9, 1.8) {$\tilde\psi_i$};
\node at (-1.7+.8,.1) {$\psi_i$};
 
\filldraw (-1.2, 0.2) circle (1pt);
\filldraw (-2.2, 1.3) circle (1pt);
 
\end{tikzpicture}\hspace{1cm}\cong \hspace{1cm}
\begin{tikzpicture}[baseline=.9cm]

\draw[very thick] (-3,0) -- (-3,2);
\draw[->,>=stealth,thick] (-1.2, 0.2) -- (-2.2, 0.7);
\draw[->,>=stealth,thick] (-1.2-2.8, 1.3) -- (-2.2-2.8, 1.8);

\node at (-3.9, 1.8) {$\tilde\psi_i$};
\node at (-1.7+.8,.1) {$\psi_i$};
 
\filldraw (-1.2, 0.2) circle (1pt);
\filldraw (-2.2-1-.8, 1.3) circle (1pt);
 
\end{tikzpicture}
\end{equation}

The fact that the original boundary is conformal means that, in the unfolded theory, it defines a topological  interface. Importantly, such topological interfaces always preserve the scaling dimension of fields~\cite{Chang:2018iay}.\footnote{The argument in~\cite{Chang:2018iay} concerns topological lines but the same argument holds for topological interfaces.} Then, in our folded theory, this implies that if the in-going state has conformal weights $(h,0)$, then the outgoing state has conformal weights $(0,h)$. In other words, conformal boundaries in such theories preserve the scaling dimension and turn left-movers into right-movers.
For boundary conditions of the form~\eqref{eq:bc_bosonized} (i.e., $X\longrightarrow\mathcal R\tilde X$) conservation of the scaling dimension is manifest, since $\mathcal R$ is orthogonal. In the fermionic language (see e.g.~\eqref{eq:out_state_345}), it is less obvious; we leave a more detailed discussion to appendix~\ref{app:scaling_ferm}.

Let us now only assume the $U(1)$ symmetry and impose conservation of the scaling dimension, as per the above argument.\footnote{In actuality, we here require one further property of the scattering process, namely that the outgoing radiation is describable by exponentials of the $X,\tilde X $ fields with fractional coefficients, which is not guaranteed in general. For some boundary states it could well be that, for instance, the endpoints of non-invertible lines could excite outgoing states. For the 3450 model the assumption is justified, see appendix~\ref{app:scaling_ferm}.}
If we throw $e^{iX_1}$ at the boundary, then the most general scattering process compatible with $U(1)$ conservation takes the form of 
\begin{equation}
\ba 
e^{iX_1}\longrightarrow e^{i\frac35\tilde X_1+i\alpha\tilde X_2}
\ea 
\end{equation}
with $\alpha$ some rational coefficient, which is undetermined since $\tilde X_2$ is neutral under the $U(1)$. 
Conservation of the scaling dimension however further implies $\frac12=\frac12((3/5)^2+\alpha^2)$, so $\alpha=\pm4/5$. 
This twist operator has monodromy $3/5$ around $e^{i\tilde X_1}$ and $\pm 4/5$ around $e^{i\tilde X_2}$. On the other hand, if we send in $e^{i\tilde X_2}$, the out-state is of the form $e^{i\frac45\tilde X_1+i\beta\tilde X_2}$ with $\beta=\pm 3/5$, to conserve scaling dimension. Now, if $\sign(\alpha)=-\sign(\beta)$ we recover the process in~\eqref{eq:out_state_345} with the fermions carrying charges $0,0,3,\pm 4$ under the $\Z_5$ generator $T$ and $\tilde n_2 \rightarrow \pm \tilde n_2$. While the positive sign allows for an embedding of the discrete symmetry into $U(1)\times U(1)'$, note that the negative sign $\Z_5$ can also be embedded into a $U(1) \times U(1)''$, where the fermions have charges $4, -3, 0, -5$ under $U(1)''$.\footnote{$U(1)'$ and $U(1)''$ are essentially the same symmetry, they are related by the field redefinition $\tilde\psi_2\leftrightarrow\tilde\psi_2^\dagger$.} If, instead, $\sign(\alpha)=\sign(\beta)$ the two lines generate different symmetries which have a mixed 't Hooft anomaly. The respective outgoing fractional vertex operators will transform under the other $\Z_5$ by  $3\times 4+\alpha \beta=12(1+\sign(\alpha)\sign(\beta)) \neq 0 \mod 5$, whereas the in-state is neutral, so both discrete symmetries would be broken by such a boundary. 

That the signs of $\alpha, \beta$ are the essential ambiguity in the scattering experiment is perhaps not surprising. The Sugawara construction implies that the Virasoro generators can be expressed as a quadratic function of the generators of $U(1)\times U(1)'$. Thus, $U(1)\times\text{Vir}$ and $U(1)\times U(1)'$ contain similar constraints, with the important difference that the former leaves the sign of the $U(1)'$  charge undetermined. In fact this argument can be made precise, and shows that all conformal boundary states preserving $U(1)$ also preserve either $U(1)'$ or $U(1)''$, with the exception of a small family of pathological states that are usually ruled out on physical grounds. We leave this check to appendix~\ref{app:scaling_ferm}. Thus, ignoring these pathological states, conservation of $U(1)$ implies conservation of an additional $\Z_5$, as assumed. For completeness, we explicitly compute the $\mathbb Z_5$ charge of the out-state in appendix~\ref{app:scaling_ferm}, and show that it is indeed conserved by the scattering process.

This concludes our discussion of the scattering puzzle in the 3450 model. With an eye to the main application of this paper, namely scattering off magnetic monopoles, we make some final observations on the analogy with this toy model. As mentioned at the beginning of section~\ref{sec:u1xu1}, the system is not rational with respect to a single $U(1)$ symmetry, but we were still able to determine the symmetric boundary condition explicitly, which will usually fail for more complicated systems.

At the intuitive level, the reason a single $U(1)$ symmetry is not enough is that it only imposes \emph{one} constraint on the fermions, but we have \emph{two} fields. 
Meanwhile, as we have seen, conformal invariance provides an independent constraint and we now have as many conditions as we have fermions, making the choice of boundary state essentially unique.
More generally, if we have $N$ fermions and a symmetry $G\subseteq SO(N)$, then the symmetric boundary state is unique if $\rank(G)=N$, since in this case we have as many constraints as we have degrees of freedom. If $\rank(G)=N-1$, we still are able to classify Cardy states thanks to conformal invariance. In more technical terms, if $\rank(G)=N$ then the embedding $G\subseteq SO(N)_1$ is conformal and the boundary state is unique, while if $\rank(G)=N-1$ the coset $SO(N)_1/G$ has $c=1$, whose boundary states are still under control. If $\rank(G)<N-1$, the classification of Cardy states really becomes unfeasible, and the out-state cannot be uniquely determined by symmetry considerations alone. This includes, for example, a version of the 3450 model containing three left-movers and three right-movers. As we shall see, the monopole scattering problem also involves, generically, too many fermions to be able to entirely fix the out-state using only symmetries, and we will have to make some further assumptions.

\section{Monopole Scattering in $\boldsymbol{4d}$}\label{sec:non-chiral_4d}

In this section we consider the scattering of light charged fermions off heavy monopoles in $4d$. We will find that many ideas are very similar to the $2d$ toy model from the previous section. Specifically, the non-trivial aspect of this scattering experiment will stem from an apparent impossibility in writing down an out-state consistent with the conservation laws, which will force us to include states with fractional occupation numbers. As before, these must be interpreted as regular excitations in a twisted sector, i.e., the out-state will be attached to some topological defect. We begin by reviewing some known facts about $4d$ monopoles and how, at low-energies, and to leading order in the fine structure constant, the scattering process becomes effectively a problem concerning massless two-dimensional fermions. At the end of this section we discuss some corrections that appear at higher orders in $e^2$, in particular, the appearance of a small centrifugal barrier at sub-leading orders in $e^2$ for certain modes.

\subsection{Massless Fermions in a Monopole Background}

Magnetic monopoles do not occur  as finite-energy solutions in $4d$ abelian gauge theories, but they can be put into the theory by hand as singular background configurations known as Dirac monopoles
\be\label{Dmonopole}
A^\text{Dirac}=-g(1-\cos \theta) \mathrm d\phi\,.
\ee
These should be regarded as line defects ('t Hooft lines) in these abelian gauge theories. These line defects are charged under the magnetic $1$-form symmetry and hence the magnetic charge is not screened. Such 't Hooft lines can arise from (heavy) finite energy excitations in a non-abelian Yang-Mills theory that is Higgsed to some subgroup containing abelian factors (see e.g.~\cite{Shnir:2005vvi,Tong:2018gt} for reviews). The simplest version of this is the $SU(2)$ Georgi-Glashow model~\cite{PhysRevD.6.2977} breaking down to its $U(1)$ torus.

Here we take an effective field theory approach, and consider a $4d$ gauge theory consisting of a $U(1)$ gauge field coupled to $N_f$ Dirac fermions $\Psi_i$ with unit charge, in the presence of a monopole of magnetic charge $n>0$, where the charge is quantized as $eg=2\pi n$. For the most part, we will be agnostic to concrete UV completions of the theory, given that the interesting physics happens in the IR.

Below we work in the tree level approximation, ignoring gauge field fluctuations around the background~\eqref{Dmonopole}. The Dirac equation in a static monopole background $A^\text{Dirac}$ commutes with the total angular momentum operator
\be
\label{eq:conserved_SU2_J}
\boldsymbol J= \boldsymbol r \times(\boldsymbol p-\boldsymbol A)+\half \boldsymbol {\sigma}-\frac{n}{2}\hat{\boldsymbol{r}}\,,
\ee 
which has eigenfunctions
\be 
\ba 
\boldsymbol J^2 \phi^{(\lambda)}_{j,\mu}(\Omega)&=j(j+1)\phi^{(\lambda)}_{j,\mu}(\Omega)\,,\\
J_z \phi^{(\lambda)}_{j,\mu}(\Omega)&=\mu \phi^{(\lambda)}_{j,\mu}(\Omega)\,,
\ea 
\ee 
where 
\be
\ba 
j&=j_0, j_0+1,\dots\,, \qquad \text{with} \quad j_0=\frac{n-1}{2}\,,\\
\mu&=-j,-j+1,\dots,j\,.
\ea
\ee
Here $\Omega$ represents the angular coordinates $\theta, \phi$ on a sphere surrounding the monopole, and $\lambda=1,2$ labels the degenerate eigenfunctions when $j\neq j_0$. $\phi^{(\lambda)}_{j,\mu}$ are 2-vectors constructed in \cite{WU1976365} in terms of monopole spherical harmonics. For $j=j_0$ there is a unique eigenfunction, namely $\phi^{(2)}_{j_0,\mu}$. The spin-$j$ solution to the Dirac equation with energy $E$ was found in \cite{PhysRevD.15.2287}. Interpreting the $j$-mode as a wave function, one straightforwardly finds from the explicit form of the solution that the probability to encounter the fermion near the core of the monopole $r \sim 0$ is 
\be\label{eq:centrifuge}
r^2\vert \Psi_{i,j,E}\vert^2 \propto r^{2\nu}\,, \qquad \nu=\sqrt{(j-j_0)(j+j_0+1)}\,.
\ee
Clearly, only the lowest angular momentum mode $j=j_0$ reaches the core of the monopole and can interact with the degrees of freedom trapped there. Hence, it is suggestive\footnote{To leading order in $e^2$ the angular momentum modes are decoupled and ignoring the higher angular momentum modes is rigorously justified. Some corrections due to finite $e^2$ are discussed later.} to forget about the modes with $j>j_0$, since they do not see the monopole. For the unit charge monopole, where $j_0=0$, this is the so-called $s$-wave approximation. The time-dependent solution for the $j=j_0$ mode is
\be
\label{eq:j0ferm}
\Psi_{i,j_0}=\sum_\mu \frac{1}{r}
\begin{pmatrix}
	\psi_{i,\mu}(t+r) \, \phi^{(2)}_{j_0,\mu} (\Omega) \\
	\tilde \psi^\dagger_{i,\mu} (t-r) \, \phi^{(2)}_{j_0,\mu} (\Omega)
\end{pmatrix}\,,
\ee
which effectively reduces the scattering off the monopole to a problem of $nN_f$ left-moving ($\psi_{i,\mu}$) and right-moving ($\tilde \psi_{i,\mu}$) $2d$ fermions. The entirety of the data of the UV and the monopole is encoded in some non-trivial boundary condition at $r=0$. An important feature of the $j_0$-mode is that the chirality and whether the mode is in-falling/outgoing are correlated.

Before moving on, there is a point that we should stress here. The modes with $j>j_0$ all come in pairs $\lambda=1,2$ of opposite helicity; in a scattering experiment, the incoming $j$-th mode evolves into the outgoing $j$-th mode, with some phase shift that is determined by the details of the experiment in the usual way. By contrast, the lowest-lying mode $j=j_0$ is unpaired, and the scattering process involves some reshuffling of these fermions. It is this reshuffling of ``unpaired zero-modes'' that gives rise to non-trivial out-states, which typically live in a different Fock sector of the theory. It is important to keep in mind that it is the monopole singularity that leads to these unpaired modes; in a regular scattering experiment where the target is not magnetically charged, all modes are typically paired up, and the outgoing radiation consists of the standard wave packets we are used to in scattering theory.

\subsection{Symmetries Preserved by the Monopole}\label{sec:resolution}

We argued above that the $4d$ scattering process can be reduced, at low energies and to leading order in $e^2$, to an effective $2d$ problem of scattering free fermions off a boundary. What symmetries of the original $4d$ system does this boundary preserve?

Four-dimensional QED has $0$-form symmetry (including the gauge group  for bookkeeping purposes, and up to discrete factors that we leave implicit for now)
\begin{equation}\label{eq:QED_IR_symm}
[U(1)_\text{gauge}]\times SU(N_f)_L\times SU(N_f)_R\times\mathcal D_{p/N}~.
\end{equation}
$\mathcal D_{p/N}$ represents the axial symmetry, which allows axial rotations by rational angles and is non-invertible due to the ABJ anomaly~\cite{Choi:2022jqy,Cordova:2022ieu}. An important fact is that, in the presence of $n$ units of magnetic flux through the sphere, only axial rotations that are generated by $e^{\frac{2\pi i}{2nN_f}}$ act on the $2d$ fermions. In other words only $\mathbb{Z}_{2nN_f}$ acts on the $2d$ fermions. The rest of the topological surfaces vanish when wrapped on $S^2$.\footnote{\label{SymmFoot}The $\mathbb{Z}_{2nN_f}$ can also be understood by elementary means as follows: from the standard ABJ anomaly, we infer that transforming the fermions with phase $e^{\frac{2\pi i}{2nN_f}}$, the $\theta$ angle shifts by $2\pi/n$. In the sector of $n$ units of magnetic flux, integrating over the $S^2$, we see that the $2d$ $\theta$ angle shifts by $2\pi$ and hence $\mathbb{Z}_{2nN_f}$ is a symmetry. More formally, the axial symmetry acting on the $2d$ fermions is broken to a discrete subgroup due to its mixed anomaly with the gauge group. Specifically, given that
\begin{equation}\label{eq:mixed_v_a}
\tr[U(1)_V\text{-}U(1)_A]=2nN_f
\end{equation}
the axial symmetry becomes $\mathbb Z_{2nN_f}$. This is just a restatement of the half-gauging construction in~\cite{Choi:2022jqy,Cordova:2022ieu}. Note that only $\mathbb Z_{2nN_f}/\mathbb{Z}_2$ is faithfully acting -- this will be important later.}  
Therefore we can think of the symmetry group of QED in the presence of $n$ units of magnetic flux through the $S^2$ as (we continue to ignore discrete identifications for now)
\begin{equation}\label{eq:QED_IR_symm_New}
[U(1)_\text{gauge}]\times SU(N_f)_L\times SU(N_f)_R\times \mathbb{Z}_{2nN_f}~.
\end{equation}

The defects that implement these symmetries do not in general admit topological junctions with the 't Hooft lines. In other words, the monopole boundary conditions do not preserve all of these symmetries. Technically, this is because when we realize the symmetries~\eqref{eq:QED_IR_symm_New} on the $2d$ modes, some of the symmetries have 't Hooft anomalies in $2d$ and hence are necessarily explicitly broken by simple boundary conditions (but not away from the boundary). 
Another way to state the condition for the $0$-form symmetry~\eqref{eq:QED_IR_symm_New} to have a topological junction with the 't Hooft line is that the $0$-form symmetry does not participate in a 2-group involving the magnetic $U(1)^{(1)}$ $1$-form symmetry.\footnote{See e.g.~\cite{Bhardwaj:2022yxj,Bhardwaj:2022lsg,Freed:2022qnc,Bhardwaj:2022kot,Bartsch:2022ytj,Bhardwaj:2023wzd,Bhardwaj:2023ayw,Bartsch:2023pzl,Bartsch:2023wvv,Copetti:2023mcq} for recent developments in obtaining a higher category theory formulation of symmetries in QFT.}
The symmetries in~\eqref{eq:QED_IR_symm_New} that participate in a 2-group would correspond to symmetries in $2d$ with 't Hooft anomalies and therefore would be explicitly broken at the $r=0$ boundary but they would be exact symmetries of the $2d$ modes for $r\neq0$. (Note that if the monopoles become dynamical, the symmetries that participate in a 2-group must be emergent in the infrared.)  

There are essentially two subgroups of $SU(N_f)_L\times SU(N_f)_R$ that have vanishing $2$-group structure constants with the $U(1)^{(1)}$. They are the two diagonal subgroups $SU(N_f)_V$ under which $\Psi_L,\Psi^c_R$ transform as $(\ydiagram1,\ydiagram1)$ and $(\ydiagram1,\overline{\ydiagram1})$, respectively. This is an elementary fact from the $2d$ point of view, as $SU(N_f)_{L/R}$ have the standard $2d$ 't Hooft anomalies, being that they only act on a single chirality. Therefore the diagonal or the anti-diagonal subgroups do not have 't Hooft anomalies. Which of these (if any) is preserved by the monopole boundary condition depends on the choice of UV completion. For example, the usual completion into an $SU(2)$ gauge theory leads to the choice $(\ydiagram1,\ydiagram1)$. 
We shall refer to these two versions of the theory as $\operatorname{QED}(\ydiagram1,\ydiagram1)$ and $\operatorname{QED}(\ydiagram1,\overline{\ydiagram1})$, respectively. Although $(\ydiagram1,\ydiagram1)$ exhibits a cubic 't Hooft anomaly for $SU(N_f)_V$ in $4d$, which would preclude a boundary~\cite{Jensen:2017eof,Thorngren:2020yht}, it presents no such obstruction to a line defect, and has no corresponding anomaly in the $2d$ reduction.

Now we have to consider the axial $\mathbb{Z}_{2nN_f}$ acting on the $2d$ fermions. This again has 't Hooft anomalies which are due to a 2-group structure in four dimensions.\footnote{\label{DiscAno}In $2d$, the 't Hooft anomaly of $\mathbb Z_{2nN_f}$ can be exhibited as follows. First, we note that $-1\in\mathbb Z_{2nN_f}$ is also in $-1\in U(1)_\text{gauge}$ so only $\mathbb Z_{nN_f}$ acts faithfully. We can remove this double-counting by composing the axial symmetry with a gauge transformation such that $\mathbb Z_{nN_f}$ acts on the left-movers alone -- this new symmetry has no overlap with the gauge group and its anomalies can be determined in a straightforward fashion. As we review in appendix~\ref{app:anomalies}, this group is anomaly-free if $nN_f$ is odd, but it has a non-perturbative anomaly if $nN_f$ is even. Hence, the subgroup of the axial symmetry that is preserved by the boundary is $\mathbb Z_{nN_f/\gcd(2,nN_f)}$.} The symmetry that is compatible with the boundary condition is therefore reduced. After accounting for discrete identifications, the subset of~\eqref{eq:QED_IR_symm} that is preserved by the $2d$ system with its simple boundary is
\begin{equation}\label{eq:QED_UV_symm}
\begin{aligned}
\operatorname{QED}(\ydiagram1,\ydiagram1) &\colon\quad \frac{[U(1)_\text{gauge}]\times SU(N_f)_V\times \mathbb Z_{nN_f}\times SU(2)_J}{\mathbb Z_2\times\mathbb Z_{\gcd(2,nN_f)}\times\mathbb Z_{N_f}}\\[1ex]
\operatorname{QED}(\ydiagram1,\overline{\ydiagram1})&\colon\quad \frac{[U(1)_\text{gauge}]\times SU(N_f)_V\times \mathbb Z_{n}\times SU(2)_J}{\mathbb Z_2\times\mathbb Z_{\gcd(2,n)}\times\mathbb Z_{N_f}}
\end{aligned}
\end{equation}
where $SU(2)_J$ denotes spacetime rotations around the monopole, generated by~\eqref{eq:conserved_SU2_J}. 
The discrete quotients are generated by the following trivially-acting elements:
\begin{equation}\label{eq:discrete_quotient}
\begin{aligned}
\mathbb Z_2&=\langle ((-1)^{n+1},1,1,-1)\rangle\\
\mathbb Z_{\gcd(2,nN_f)}\ \text{or}\ \mathbb Z_{\gcd(2,n)}&=\langle (-1,1,-1,1)\rangle\\
\mathbb Z_{N_f}&=\begin{cases}
\langle (1,e^{2\pi i/N_f},e^{-2\pi i/N_f},1)\rangle&\operatorname{QED}(\ydiagram1,\ydiagram1)\\
\langle (e^{2\pi i/N_f},e^{-2\pi i/N_f},1,1)\rangle&\operatorname{QED}(\ydiagram1,\overline{\ydiagram1})
\end{cases}
\end{aligned}
\end{equation}

The fermions transform under these symmetries according to
\begin{equation}\label{eq:monopole_1st_quant_nums}
\begin{aligned}
&\qquad\operatorname{QED}(\ydiagram1,\ydiagram1)\\[+1ex]
&\begin{array}{c|cc}
&\psi&\tilde\psi\\ \hline
[U(1)_\text{gauge}]&1&-1\\
SU(N_f)_V&\ydiagram1&\ydiagram1\\
SU(2)_J&\boldsymbol n&\boldsymbol n\\
\mathbb Z_{nN_f}&1&1
\end{array}\end{aligned}\hspace{3cm}
\begin{aligned}
&\qquad\operatorname{QED}(\ydiagram1,\overline{\ydiagram1})\\[+1ex]
&\begin{array}{c|cc}
&\psi&\tilde\psi\\ \hline
[U(1)_\text{gauge}]&1&-1\\
SU(N_f)_V&\ydiagram1&\overline{\ydiagram1}\\
SU(2)_J&\boldsymbol n&\boldsymbol n\\
\mathbb Z_{n}&1&1
\end{array}
\end{aligned}
\end{equation}
The fermion $\psi$ has $2d$ chirality $\gamma^\star=+1$, which corresponds to ingoing motion in $4d$. The fermion $\tilde\psi$ has $2d$ chirality $\gamma^\star=-1$, which corresponds to outgoing motion in $4d$. The scattering process consists of throwing $\psi$ towards the boundary, with some excitation of the field $\tilde\psi$ coming out, in such a way that the quantum numbers above are conserved.

Thus, all in all, the final conclusion is that the subgroup of the $4d$ flavor symmetry that does not participate in a 2-group precisely agrees with the subgroup of $2d$ flavor symmetry that does not have a 't Hooft anomaly. This is as expected, since this subgroup corresponds to the symmetries that are preserved by the monopole, an obstruction that is measured by a 2-group in $4d$ and by an 't Hooft anomaly in $2d$, being that the monopole is a boundary condition.

More generally, the interplay between $4d$ and $2d$ symmetries and anomalies when we compactify  on a sphere with magnetic flux is as follows (related discussions can be found in~\cite{Cordova:2018cvg,Maldacena:2020skw,Aharony:2022ntz,ZoharDraft}). Consider a $4d$ abelian gauge theory with left-handed Weyl fermions with gauge charges $q_i$. The theory is only consistent if the gauge anomaly vanishes $\sum_i q_{i}^3=0$, and can be defined on a curved background if it does not posses a gauge-gravitational anomaly, $\sum_i q_{i}=0$. (These are the gauge anomaly cancellation conditions arising from triangle diagrams, i.e., they correspond to perturbative anomalies. There are no non-perturbative anomalies for $U(1)$ --- neither in $2d$ nor in $4d$ --- since the corresponding bordism groups vanish, $\Omega_5^\text{Spin}(BU(1))=\Omega_3^\text{Spin}(BU(1))=0$, cf.~e.g.~\cite{Garcia-Etxebarria:2018ajm}).

Consider some global symmetry, and let us without loss of generality focus on its maximal torus $U(1)^r$, which acts as
\be 
\psi_{i} \mapsto e^{i \alpha_a Q_{ai}} \psi_{i}\,,
\ee
for some matrix of charges $Q_{ai}$, where $a$ runs over the $U(1)$ factors. 
A $U(1)_a\subset U(1)^r$ symmetry has an ABJ anomaly unless $\sum_i Q_{ai} q_{i}^2 = 0$. Furthermore, $\sum_i Q_{ai}^2 q_i\neq0$ implies that $U(1)_a$ mixes with the magnetic $1$-form symmetry to form a $2$-group.

Let us see how these conditions translate into statements for the $2d$ theory. Reduction on a sphere with magnetic flux $n$ gives rise to $n q_{i}$ left-moving fermions $\psi_{i,\mu}$ for $q_i>0$, as well as $-nq_{i}$ right-moving fermions $\tilde{\psi}_{i,\mu}$ for $q_i<0$. For there to exist a conformal (presumably requiring energy conservation is enough) boundary condition in $2d$, the chiral central charges must coincide $c_L=c_R$, i.e., the number of left-/right-moving fermions must be the same,
\be 
\sum_{q_i>0} n q_{i}=-\sum_{q_i<0} n q_{i}\,.
\ee
Clearly, this holds if the $4d$ theory has no gauge-gravitational anomaly. This follows from the absence of 2-group symmetry involving space-time symmetries.

 In the presence of a magnetic monopole, the condition for a $U(1)_a$ and $U(1)_b$ symmetry to be anomaly free in $2d$ is
\be 
\label{eq:2danomfree}
\sum_i n q_i \times Q_{ai} Q_{bi}=0\,.
\ee
This follows from the absence of a 2-group symmetry involving the $U(1)_a$ and $U(1)_b$ symmetries.

Gauge invariance of the 't Hooft line implies that $U(1)_\text{gauge}$ must be conserved by the boundary, which in turn is only possible if it is anomaly free in $2d$,
\be 
\sum_{i} nq_{i} \times q_{i}^2=0\,,
\ee
which holds for any consistent $4d$ gauge theory, i.e., without cubic anomalies. A similar calculation to~\eqref{eq:2danomfree} shows that a mixed anomaly in $2d$ between $U(1)_\text{gauge}$ and some $U(1)_a$ is absent if
\be
\sum_{i} nq_i \times q_i Q_{ai} = 0\,.
\ee
which is the condition for $U(1)_a$ to have no ABJ anomaly. The proper interpretation is that in $4d$ $U(1)_a$ becomes non-invertible and the $2d$ theory enjoys an action only by an invertible subgroup, which may still have some 't Hooft anomalies that reduce it further by the boundary conditions. (We worked out an example of these ideas above.)

\subsection{Resolution to the Monopole Scattering Problem}\label{sec:monopole_2d_solution}

We now have all the tools we need to discuss the monopole scattering problem. We saw above that such an experiment can be reduced, at low energies and to leading order in $e^2$, to an effective $2d$ problem consisting of free fermions $\psi_{i,\mu},\tilde\psi_{i,\mu}$ propagating on the half plane $(r,t)\in\mathbb R_+\times\mathbb R$, and that the interactions with the monopole are encapsulated in some boundary conditions at $r=0$. This is basically the same setting we discussed in section~\ref{sec:2dBCFT}. Then, the scattering process can be described as a boundary problem where the left-movers $\psi_{i,\mu}$ turn into right-movers $\tilde\psi_{i,\mu}$ in a way consistent with the symmetries of the system~\eqref{eq:monopole_1st_quant_nums}.

In general, the free fermion CFT $\{\psi,\tilde\psi\}$ is not rational with respect to the symmetry algebra preserved by the boundary. Then, as discussed in section~\ref{sec:2dBCFT}, there are infinitely many boundary states consistent with the symmetry, and we cannot classify such states efficiently. Therefore, the outgoing state is not fixed by symmetry alone, and the final answer to the scattering experiment is sensitive to the details of the UV completion, equivalently, the precise choice of boundary state. We will begin by considering the case of the minimally charged monopole $n=1$, for which the symmetry does determine the out-state uniquely, and then we will generalize the answer to higher $n$.

\paragraph{Unit charge monopole.} When $n=1$, the symmetries of the system~\eqref{eq:monopole_1st_quant_nums} simplify to
\begin{equation}
\begin{aligned}
&\qquad\operatorname{QED}(\ydiagram1,\ydiagram1)\\[+1ex]
&\begin{array}{c|cc}
&\psi&\tilde\psi\\ \hline
U(1)_\text{gauge}&1&-1\\
SU(N_f)_V&\ydiagram1&\ydiagram1
\end{array}\end{aligned}\hspace{3cm}
\begin{aligned}
&\qquad\operatorname{QED}(\ydiagram1,\overline{\ydiagram1})\\[+1ex]
&\begin{array}{c|cc}
&\psi&\tilde\psi\\ \hline
U(1)_\text{gauge}&1&-1\\
SU(N_f)_V&\ydiagram1&\overline{\ydiagram1}
\end{array}
\end{aligned}
\end{equation}

In $\operatorname{QED}(\ydiagram1,\overline{\ydiagram1})$, the symmetry-preserving boundary condition is easy to write down,
\begin{equation}\label{eq:bar_box_n_1}
\psi\bigl|_{\mathcal B}=e^{i\theta}\tilde\psi^\dagger\bigl|_{\mathcal B}\,,
\end{equation}
which implies that the scattering process is
\begin{equation}\label{eq:bar_box_n_1_arrow}
\psi\longrightarrow \tilde\psi^\dagger\,,
\end{equation}
This manifestly conserves $U(1)_\text{gauge}\times SU(N_f)_V$. The parameter $\theta$ is often interpreted as a  marginal parameter on the 't Hooft line, see~\cite{Yamagishi:1982wp,Aharony:2022ntz}.

On the other hand, in $\operatorname{QED}(\ydiagram1,\ydiagram1)$, there is no obvious candidate for the out-state that conserves the symmetries. A naive scattering amplitude $\psi\longrightarrow \tilde\psi$ does not work, since it does not conserve $U(1)_\text{gauge}$, while an amplitude $\psi\longrightarrow \tilde\psi^\dagger$ (which does preserve electric charge) is incompatible with $SU(N_f)_V$ invariance (unless $N_f\le 2$, given that $\ydiagram1\cong\overline{\ydiagram1}$ in $SU(2)$). This is known as the Callan-Rubakov problem~\cite{Rubakov:1982fp,Callan:1982ah,Callan:1982ac}. The resolution is --- as in the 3450 model --- that the outgoing state lives in a twisted sector. We will see that in a certain twisted Hilbert space, a ``single-particle'' excitation exists with the correct quantum numbers. (``Single-particle'' is in quotation marks because particle number is not conserved in the scattering process.)

We can look for the required twist operator either in the fermionic picture or in the bosonic one. In this section, the bosonized dual is perhaps less useful than in the 3450 model, since here the symmetry group is non-abelian (and hence we would need to use non-abelian bosonization~\cite{Witten:1983ar}, which is clunkier than simply working with our free fermions $\psi,\tilde\psi$). A nice compromise is to forget about the full symmetry $[U(1)_\text{gauge}]\times SU(N_f)_V$, and focus on the maximal torus $U(1)^{N_f}$. If we do so, we can perform the standard bosonization map $\psi\leftrightarrow e^{iX}$, and write down an outgoing vertex operator that conserves the required symmetries. The resulting out-state will \emph{manifestly} conserve the Cartan, but the conservation of the full group will not be obvious, and requires a separate check. (This check is carried out in \cite{Smith:2020nuf}.)

Let us proceed with abelian bosonization. If we write $\psi_i\leftrightarrow e^{iX_i}$, $\tilde\psi_i\leftrightarrow e^{i\tilde X_i}$, then $[U(1)_\text{gauge}]$ acts as
\begin{equation}
[U(1)_\text{gauge}]\colon\quad X_i\mapsto X_i+2\pi \alpha\,,
\end{equation}
while $U(1)^{N_f-1}\subset SU(N_f)_V$ acts as
\begin{equation}
U(1)^{N_f-1}\colon\quad X_i\mapsto X_i+2\pi \alpha_i,\qquad X_{i+1}\mapsto X_{i+1}-2\pi \alpha_i\,.
\end{equation}
We can write this more compactly as
\begin{equation}
U(1)^{N_f}\colon\quad\begin{aligned}X_i\mapsto X_i+2\pi Q_{ij}\alpha_j\\\tilde X_i\mapsto \tilde X_i+2\pi \tilde Q_{ij}\alpha_j\end{aligned}
\end{equation}
where
\begin{equation}
\begin{aligned}
&\operatorname{QED}(\ydiagram1,\ydiagram1)\colon\qquad 
 Q=\left( \begin{smallmatrix}
 	+ & + & + & \cdots & +\\
 	+ & - & & & &\\
 	 & + & - & & &\\
 	 & & & \ddots & &\\
 	 & & & & - &
 \end{smallmatrix}\right) \,,\qquad
 \tilde Q=\left( \begin{smallmatrix}
 	- & - & - & \cdots & -\\
 	+ & - & & & &\\
 	 & + & - & & &\\
 	 & & & \ddots & &\\
 	 & & & & - &
 \end{smallmatrix}\right)\\
&\operatorname{QED}(\ydiagram1,\overline{\ydiagram1})\colon\qquad 
 Q=\left( \begin{smallmatrix}
 	+ & + & + & \cdots & +\\
 	+ & - & & & &\\
 	 & + & - & & &\\
 	 & & & \ddots & &\\
 	 & & & & - &
 \end{smallmatrix}\right) \,,\qquad
 \tilde Q=\left( \begin{smallmatrix}
 	- & - & - & \cdots & -\\
 	- & + & & & &\\
 	 & - & + & & &\\
 	 & & & \ddots & &\\
 	 & & & & + &
 \end{smallmatrix}\right)
 \end{aligned}
 \end{equation}

The boundary state preserving this $U(1)^{N_f}$ symmetry is commonly known as the \emph{dyon boundary condition} in the literature (see e.g.~\cite{Affleck:1993np,Smith:2020nuf}). The conservation of this group determines the out-state to be
\begin{equation}
X_i\longrightarrow \mathcal R_{ij}\tilde X_j\,,
\end{equation}
where the matrix $\mathcal R:=Q^{-1}\tilde Q$ is given by
\be 
\label{eq:dyonbdycond}
\begin{aligned}
\operatorname{QED}(\ydiagram1,\ydiagram1)\colon\qquad \cR&=\mathds{1}-\frac{2}{N_f}\\
\operatorname{QED}(\ydiagram1,\overline{\ydiagram1})\colon\qquad \cR&=-\mathds{1}
\end{aligned}
\ee

For $\operatorname{QED}(\ydiagram1,\overline{\ydiagram1})$, this means that the scattering process takes the form
\begin{equation}\label{eq:n_1_QED2}
e^{iX_i}\longrightarrow e^{-i\tilde X_i}\,,
\end{equation}
which is just the bosonic version of~\eqref{eq:bar_box_n_1_arrow}.\footnote{In the bosonic language the $\theta$ angle in~\eqref{eq:bar_box_n_1_arrow} simply corresponds to working with a more general class of boundary conditions $X=\mathcal R\tilde X+\theta$ as mentioned below~\eqref{eq:bc_bosonized}.} On the other hand, for $\operatorname{QED}(\ydiagram1,\ydiagram1)$, the scattering process becomes
\be
\label{eq:monopole_n_1_bosonic_out}
e^{iX_i} \longrightarrow e^{ i\tilde X_i-i\frac{2}{N_f}\sum_j\tilde X_j }\,,
\ee
so in this case the outgoing radiation contains fractionalized states.

The analogy with the 3450 model is clear; instead of associating this state with fractional occupation numbers for the fermions, we interpret it as the appearance of a topological line. We can split the twist operator into a local piece and the vacuum of the twisted sector 
\be 
e^{ i\tilde X_i-i\frac{2}{N_f}\sum_j\tilde X_j }=\prod_j e^{in_{ij}\tilde X_j}\times e^{i (\delta_{ij}-\frac{2}{N_f}-n_{ij})\tilde X_j }\,,
\ee 
with $n_{ij}\in \Z$, such that the vacuum has the smallest possible scaling dimension, implying $n_{ij}=\delta_{ij}$ for $N_f>4$. 

The presence of the line is evidenced by the twist operator having non-trivial monodromy with the local fields of the theory. Specifically, the out-state in~\eqref{eq:monopole_n_1_bosonic_out} is local to left-movers, and has monodromy $-2/N_f$ around the right-movers. Therefore, the topological line implements a $\mathbb Z_{N_f}$ symmetry under which the left-movers are uncharged while the right-movers carry charge $-2$. 

If we let $T$ denote the generator of $\mathbb Z_{N_f}$, then the fermionic translation of~\eqref{eq:monopole_n_1_bosonic_out} is
\begin{equation}\label{eq:monopole_n_1_fermonic_out}
\psi_i\longrightarrow T^{-2}+\tilde\psi_i\,.
\end{equation}
The interpretation of this result is that, if we throw the left-moving fermion $\psi$ towards the boundary, what comes out is the right-moving fermion $\tilde\psi$ attached to a topological line that acts as $\tilde\psi\mapsto e^{2\pi i(-2/N_f)}\tilde\psi$. Note that this $\mathbb Z_{N_f}$ differs from the central $\mathbb Z_{N_f}=Z(SU(N_f))$ by a gauge transformation, so it is essentially the same symmetry.

By construction, the out-state in~\eqref{eq:monopole_n_1_fermonic_out} preserves $U(1)^{N_f}$, although this might not be entirely obvious in the fermionic language. Let us check that the symmetry is indeed conserved. If we use our general formula~\eqref{eq:general_Qvac}, we find that $T^k$ adds $Q_\text{vac}=N_f(\frac12-[\frac12+\frac{k}{N_f}])$ units of $U(1)_\text{gauge}$ charge, which equals $+2$ for $k=-2$. Therefore, the out-state carries $Q_\text{gauge}(T^{-2}+\tilde\psi)=+1$, matching the gauge charge of the in-state $\psi$. In the fermionic language, the full non-abelian symmetry $SU(N_f)_V$ is matched in a manifest way. Indeed, a line that commutes with a semi-simple group cannot add charge under this group,\footnote{Ultimately this follows from the fact that there are no $0+1d$ Chern-Simons terms for semi-simple groups that we could stack on the line. More concretely, if we use~\eqref{eq:general_Qvac} to compute the charge of $T$ under an arbitrary $U(1)\in SU(N_f)$, we find $Q_\text{vac}\propto \tr(t)$, with $t$ the generator of this $U(1)$. Given that all generators in a semi-simple group are traceless, the line $T$ does not add any charge under it, and thus its endpoint is always neutral under semi-simple groups. We stress that this conclusion holds only under the assumption that the twisted Hilbert space has no zero-modes, since these can modify the equation~\eqref{eq:general_Qvac}. Zero-modes can and often do transform non-trivially under symmetry groups, whether semi-simple or not.\label{fn:semi-simple}} and therefore $T^{-2}+\tilde\psi$ transforms in the same $SU(N_f)_V$ representation as $\tilde\psi$ itself, to wit, the fundamental representation, matching the $SU(N_f)_V$ charge of the in-state $\psi$.

The conclusion of the discussion so far is that the scattering process involving the minimally charged monopole $n=1$ takes the form
\begin{equation}
\begin{alignedat}{2}
&\operatorname{QED}(\ydiagram1,\ydiagram1)\colon\qquad &\psi_i&\longrightarrow T^{-2}+\tilde\psi_i\\
&\operatorname{QED}(\ydiagram1,\overline{\ydiagram1})\colon\qquad &\psi_i&\longrightarrow \tilde\psi_i^\dagger
\end{alignedat}
\end{equation}
where $T^{-2}$ is (up to a gauge transformation) the generator of $\mathbb Z_{N_f}\subset SU(N_f)_V$. Our next task is to generalize this to higher $n$.

\paragraph{Higher charge monopole.} We will now extend the previous discussion to $n>1$. The goal is to find an out-state, involving $\tilde\psi$, that matches the quantum numbers of $\psi$. The charges of these fields under the symmetries of the problem are given by~\eqref{eq:monopole_1st_quant_nums}, which we reproduce again here for convenience:
\begin{equation}\label{eq:repeat_eq}
\begin{aligned}
&\qquad\operatorname{QED}(\ydiagram1,\ydiagram1)\\[+1ex]
&\begin{array}{c|cc}
&\psi&\tilde\psi\\ \hline
U(1)_\text{gauge}&1&-1\\
SU(N_f)_V&\ydiagram1&\ydiagram1\\
SU(2)_J&\boldsymbol n&\boldsymbol n\\
\mathbb Z_{nN_f}&1&1
\end{array}\end{aligned}\hspace{3cm}
\begin{aligned}
&\qquad\operatorname{QED}(\ydiagram1,\overline{\ydiagram1})\\[+1ex]
&\begin{array}{c|cc}
&\psi&\tilde\psi\\ \hline
U(1)_\text{gauge}&1&-1\\
SU(N_f)_V&\ydiagram1&\overline{\ydiagram1}\\
SU(2)_J&\boldsymbol n&\boldsymbol n\\
\mathbb Z_{n}&1&1
\end{array}
\end{aligned}
\end{equation}

For either version of QED, there is now no local operator constructed out of $\tilde\psi$ that carries the same quantum numbers as $\psi$, so the out-state must again be a twist operator, i.e., it will consist of some number of $\tilde\psi$ modes attached to some topological line. For $n>1$, the free fermion CFT is not rational with respect to the chiral algebra preserved by the boundary, and hence we do not have total control over the boundary state. As far as our IR effective description is concerned, there are infinitely many conformal boundary states preserving the symmetries of the problem~\eqref{eq:repeat_eq}, and we cannot classify all Cardy states efficiently. Therefore, the out-state is not fixed by symmetry considerations alone, and we need to make some educated guesses. Inspired by the $n=1$ case, we propose that the out-state is again some (possibly fractional) vertex operator. Although this assumption need not hold for generic boundary conditions, it is certainly the simplest out-state one could write down. As we shall see presently, there is a unique reasonable candidate for the outgoing vertex operator, and it satisfies several consistency checks, making it the most natural guess for the scattering process.

Let us begin with $\operatorname{QED}(\ydiagram1,\ydiagram1)$. Consider some line that commutes with $SU(N_f)_V\times SU(2)_J$; given that such a line cannot add charge under semi-simple groups (cf.~footnote~\ref{fn:semi-simple}), the only vertex operator that preserves $SU(N_f)_V\times SU(2)_J$ is
\begin{equation}\label{eq:higher_n_ansatz}
\psi_{i,\mu}\longrightarrow \tilde\psi_{i,\mu}+\text{some line}\,,
\end{equation}
where the line does not depend on $i,\mu$.

We will identify the topological line in the ansatz~\eqref{eq:higher_n_ansatz} via the following constraints:
\begin{itemize}
\item The line has to end topologically on the boundary. This means that it has to generate some subset of the groups in~\eqref{eq:repeat_eq}, since these are by definition the symmetries preserved by the monopole.
\item The endpoint of the line must be purely right-moving, i.e., it should not carry momentum in the left-moving direction. This means that the line has to act on the right-movers only.
\item The endpoint of the line must carry $[U(1)_\text{gauge}]$ charge so as to conserve electric charge in~\eqref{eq:higher_n_ansatz}. Specifically, the field $\psi$ has charge $+1$ and $\tilde\psi$ has charge $-1$, so we require the line to carry charge $+2$ units of charge. 
\end{itemize}
These conditions are enough to uniquely fix the line. If we compose the chiral $\mathbb Z_{nN_f}$ with a gauge transformation, we can obtain a new $\mathbb Z_{nN_f}$ that only acts on the right-movers. Let $T$ be the generator of this symmetry, such that $T\colon\tilde\psi\mapsto e^{2\pi i/nN_f}\tilde\psi$. Then, using our general formula~\eqref{eq:general_Qvac}, we find that $T^k$ adds $Q_\text{vac}$ units of $[U(1)_\text{gauge}]$ charge, where
\begin{equation}
\begin{aligned}
Q_\text{vac}&=\sum_\text{right}\bigl(\tfrac12-[\tfrac12+\tfrac{k}{nN_f}]\bigr)\\
&=nN_f\bigl(\tfrac12-[\tfrac12+\tfrac{k}{nN_f}]\bigr)\,.
\end{aligned}
\end{equation}
Therefore, $T^{-2}$ adds $+2$ units of electric charge, and the general scattering amplitude becomes
\begin{equation}\label{eq:monopole_higher_n_fermonic_out}
\psi_{i,\mu}\longrightarrow T^{-2}+\tilde\psi_{i,\mu}\,,
\end{equation}
which conserves all the symmetries of the problem. This of course reduces to~\eqref{eq:monopole_n_1_fermonic_out} when $n=1$. In the bosonic language,~\eqref{eq:monopole_higher_n_fermonic_out} can be written as
\begin{equation}\label{eq:out-state_n}
e^{i X_{i,\mu}}\longrightarrow e^{i\tilde X_{i,\mu}-\frac{2}{nN_f}i\sum_{j,\mu'}\tilde X_{j,\mu'}}\,,
\end{equation}
which, as can easily be checked, preserves the maximal torus of $[U(1)_\text{gauge}]\times SU(N_f)_V\times SU(2)_J$.

Let us pause and make a few important remarks regarding the key result~\eqref{eq:monopole_higher_n_fermonic_out}. First, notice that the topological defect attached to the out-state is $T^{-2}$. If $nN_f$ is odd, this line generates $\mathbb Z_{nN_f}$ but, if $nN_f$ is even, it only generates a $\mathbb Z_{nN_f/2}$ subgroup. This is consistent with the symmetry structure of the theory, since the symmetry that is actually preserved by the monopole is $\mathbb Z_{nN_f/\gcd(2,nN_f)}$ (cf.~\eqref{eq:QED_UV_symm}). Note also that the process conserves the conformal symmetry, namely the out-state has spin $s=1/2$, just like the in-state. This is easily checked in the bosonic presentation~\eqref{eq:out-state_n}, viz.
\begin{equation}
s=\frac12\biggl(\bigl(1-\tfrac{2}{nN_f}\bigr)^2+(nN_f-1)\bigl(-\tfrac{2}{nN_f}\bigr)^2\biggr)\equiv\tfrac12\,.
\end{equation}
One can also show the same directly in the fermionic picture, which we leave to appendix~\ref{app:scaling_ferm}.

Finally, let us mention that the scattering process also conserves the charge associated to $\mathbb Z_{nN_f}$. This in fact is not entirely trivial since the line is local to the left-movers but not to right-movers, i.e., the in-state is neutral but the outgoing fermions are typically charged. The important point is that the endpoint of the line carries charge as well, which nicely cancels the charge of $\tilde\psi_{i,\mu}$. We show this explicitly in appendix~\ref{app:scaling_ferm}, but we can offer a more conceptual argument as follows.

\begin{figure}[h!]
\centering
\begin{tikzpicture}[baseline=0cm]

\node[scale=.8] at (4.2,1) {$g_1$};\node[scale=.8] at (4.2,-1.35) {$g_2$};

\draw (4,0) -- (4, 1.2);
\filldraw (4,0) circle (1.5pt) node[anchor=west,scale=.8] {$V_1$};
\draw (4,-.6) -- (4, -1.6);
\filldraw (4,-.6) circle (1.5pt) node[anchor=west,scale=.8] {$V_2$};

\node at (5,-.25) {$=$};
\begin{scope}[shift={(2.4,0)}]
\node[scale=.8] at (4.2,1) {$g_1$};\node[scale=.8] at (4.2,-1.35) {$g_2$};
\draw (4,0) -- (4, 1.2);
\filldraw (4,0) circle (1.5pt) node[anchor=west,scale=.8] {$V_1$};

\filldraw (4,-.6) circle (1.5pt) node[anchor=west,scale=.8] {$V_2$};
\draw (4, -1.6) to[in=-45,out=90] (3.9,-1.2) to[in=-70-10,out=90+45] (3.3,-.4) to[in=180,out=100] (4,.4) to[in=90,out=0] (4.7,-.2) to[in=90-70,out=270] (3.8,-.3) to[in=90,out=270-70] (3.6,-.6) to[in=180+45,out=-90] (3.95,-.8) to[in=-90,out=45] (4,-.6);
\node at (6,-.25) {$\times\ \langle V_1,V_2\rangle$};
\node at (7.4,-.25) {$=$};
\begin{scope}[shift={(4.8,0)}]
\node[scale=.8] at (4.2,1) {$g_1$};\node[scale=.8] at (4.2,-1.35) {$g_2$};
\draw (4,0) -- (4, 1.2);
\filldraw (4,0) circle (1.5pt) node[anchor=west,scale=.8] {$V_1$};

\filldraw (4,-.6) circle (1.5pt) node[anchor=west,scale=.8] {$V_2$};
\draw (4, -1.6) -- (4,-.6);
\node at (6,-.25) {$\times\ \langle V_1,V_2\rangle$};
\draw  (3.3,-.4) to[in=180,out=100] (4,.4) to[in=90,out=0] (4.7,-.2) to[in=90-70,out=270] (3.8,-.3) to[in=-80,out=200] (3.3,-.4);

\end{scope}
\end{scope}
\end{tikzpicture}
\caption{We consider two twist operators $V_1,V_2$ that live at the end of two invertible lines $g_1,g_2$, assumed anomaly-free. First, we move $g_2$ across $V_1$, picking up a factor of $\langle V_1,V_2\rangle\in U(1)$, which denotes the monodromy of $V_2$ around $V_1$ (i.e., the discontinuity across the branch cut). Then, we apply an $F$-move to recombine the lines. The end result is a $g_2$-line wrapping $V_1$, which gives by definition the charge of this operator under the line.}
\label{fig:Z5_charge_monodromy}
\end{figure}
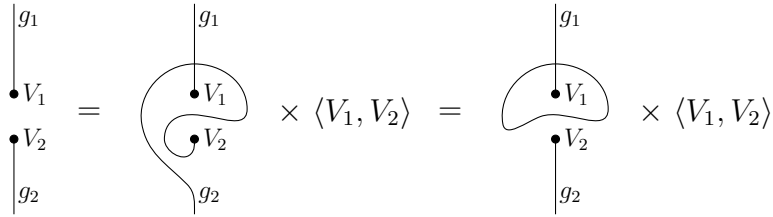

A simple way to determine the $\mathbb Z_{nN_f}$ charge of an operator $V_1$ is to consider an open $\mathbb Z_{nN_f}$ line with some twist operator $V_2$ sitting at the end. Then, the $\mathbb Z_{nN_f}$ charge of $V_1$ is given by the monodromy of $V_2$ around $V_1$ (see figure~\ref{fig:Z5_charge_monodromy}). We can apply this, for example, to $V_1=V_2=V_\text{out}$, with $V_\text{out}$ being the out-state of the scattering process. Then, this shows that the $\mathbb Z_{nN_f}$ charge of the out-state is simply its self-monodromy. But this self-monodromy manifestly vanishes, since the scattering process conserves spin (and the in-state has half-integral spin). In conclusion, as long as spin is conserved and the attached line is anomaly-free, the scattering process automatically conserves the charge measured by the outgoing line.

We now turn our attention to $\operatorname{QED}(\ydiagram1,\overline{\ydiagram1})$. Unlike in the other version of QED, here there is no line that commutes with $SU(N_f)_V\times SU(2)_J$ that is able to add the necessary charge to preserve the symmetries. The reason is that $\mathbb Z_n$ has $Q_\text{vac}\propto N_f$, and therefore it adds too much gauge charge. Consequently, we are forced to look for more exotic lines, for example a topological line that does not necessarily commute with $SU(N_f)_V$.\footnote{Other alternatives are non-invertible lines, or invertible lines that carry fermionic zero-modes. We do not explore these options here.} (The physics of the original $4d$ problem suggests that we still want the line to commute with $SU(2)_J$). After some trial and error, the following candidate emerges as the simplest ansatz:
\begin{equation}\label{eq:out_higher_n_QED2}
\psi_{i,\mu}\longrightarrow T_i+\tilde\psi_{i,\mu}\,,
\end{equation}
where $T_i$ is some topological line that only acts on the $i$-th right-moving fermion.

In order to identify $T_i$ we can play the same game as before. Such a line must generate some subgroup of $U(N_f)\times \mathbb Z_n$, since these are the symmetries of the system, and the line must also add suitable charge under such symmetries so as to obey all conservation laws. The only subgroup of $U(N_f)\times \mathbb Z_n$ that acts on the $i$-th right-moving fermion alone consists of discrete rotations
\begin{equation}
T_i\colon \psi_{j,\mu}\mapsto \psi_{j,\mu},\qquad\tilde\psi_{j,\mu}\mapsto e^{-2\pi i \delta_{ij}2/n}\tilde\psi_{j,\mu}\,.
\end{equation}
Using the by-now familiar formula~\eqref{eq:general_Qvac} we find that $T_i$ adds charge to the operators sitting at its endpoint. Concretely, consider the $U(1)_j\subset U(N_f)$ subgroup that acts on the $i$-th fermion with charges $Q_{ij}=-\tilde Q_{ij}=\delta_{ij}$; then, $T_i$ adds $Q_\text{vac}$ charge under $U(1)_j$, where
\begin{equation}
Q_\text{vac}=\delta_{ij}\sum_{\mu}(1/2-[1/2-2/n])\equiv +2\delta_{ij}
\end{equation}
(where the last equality follows for $n>3$). This way we see that $U(1)_j$ is conserved, since $\psi_{i,\mu}$ carries charge $\delta_{ij}$ and $\tilde\psi_{i,\mu}$ carries charge $-\delta_{ij}$.

In the bosonic language,~\eqref{eq:out_higher_n_QED2} can be written as
\begin{equation}
e^{iX_{i,\mu}}\longrightarrow e^{i\tilde X_{i,\mu}-\frac{2}{n}\sum_{\mu'}\tilde X_{i,\mu'}}\,.
\end{equation}
Of course this reduces to~\eqref{eq:n_1_QED2} for $n=1$. One can perform the same consistency checks as before; for example, it is easily checked that this out-state conserves scaling dimension and has vanishing self-monodromy.

The final conclusion is that, to leading order in $e^2$, the scattering process maps an incoming $j$-mode to an outgoing one, with some phase shift, except the $j=j_0$ mode, which carries a topological line and no phase shift. This defect line acts as a discrete chiral rotation (composed with a flavor rotation in $\operatorname{QED}(\ydiagram1,\overline{\ydiagram1})$). 
In section~\ref{sec:uplift_4d} we will clarify the implications of this result for our original $4d$ problem. For definiteness we will focus on $\operatorname{QED}(\ydiagram1,\ydiagram1)$, where the line acts as $\tilde\psi\mapsto e^{-4\pi i/nN_f}\tilde\psi$, although the discussion for $\operatorname{QED}(\ydiagram1,\overline{\ydiagram1})$ is rather similar.

\subsection{Effects of Gauge Field Fluctuations}\label{sec:gauged}

As long as we have not considered the gauge field fluctuations the scattering states included free $2d$ fermions moving in the space $\{r,t\}$ with boundary at $r=0$. 

Now let us consider the modes of the gauge field on $S^2$ surrounding the monopole. The components with indices tangent to the $S^2$ do not have zero modes and hence encounter a centrifugal barrier similar to that of the non-minimal spin modes of the fermions.

The $A_r$ and $A_t$ components of the gauge field are scalars on the $S^2$ and have zero modes. They have a kinetic term
$\frac{4\pi r^2}{2e^2} F_{rt}^2 $.   
Therefore the coupling to these $2d$ zero modes of the gauge field becomes strong near the monopole and weak otherwise.

Due to the coupling to the gauge field zero modes, the $U(1)_A$ symmetry  is broken to $\mathbb{Z}_{2n}$ by the $2d$ anomaly: 
$\nabla\cdot j_A \sim 2n F_{rt}$. This is a restatement of the facts explained in footnote~\ref{SymmFoot} and equation~\eqref{eq:QED_IR_symm_New}. It will be important to remember that only $\mathbb{Z}_{2n}/\mathbb{Z}_2$ is faithfully acting.
The $\mathbb{Z}_{2n}/\mathbb{Z}_2$ can be further broken at $r=0$ by the boundary condition due to 't Hooft anomalies, as we will see.

To analyze the effect of these $2d$ modes, we will first consider massless QED$_4$ with one Dirac fermion and $n=1$. The scattering states in the $s$-wave then include one complex left-moving and one complex right-moving fermion, of charges $(1,-1)$ under the gauge symmetry, respectively. We bosonize these fermions, which makes studying the effects of the gauge fields easier. We obtain a variant of the Schwinger model
\begin{equation}\label{minn} S = \int_{r=0}^\infty dr\int dt\left(\frac{1}{8\pi } (\partial \phi)^2+\frac{2\pi r^2}{e^2} F_{rt}^2  +\frac{1}{2\pi } \phi F_{rt}\right)~. \end{equation} 
 The compact boson $\phi\simeq \phi+2\pi$ has radius $R=\frac{1}{\sqrt 2}R_\text{s.d.}$ with $R_\text{s.d.}$ the self-dual radius.\footnote{For reference recall that the compact boson at radius $R$ with Lagrangian $\frac{R^2}{2}(\partial\phi)^2 $ has local operators labeled by integer $n,m$ with scaling dimensions
$$\Delta=\frac{1}{4\pi R^2}n^2+\pi R^2m^2~.$$
The boson in~\eqref{minn} has $R^2=1/4\pi$ which corresponds to the free fermion point. The original fermion, which is not a true local operator in the bosonized description, is given $n=\pm 1/2,m=1$ and the complex conjugate operators. The dual shift symmetry is gauged and corresponds to the quantum number $m$ while the axial symmetry corresponds to the quantum number $n$ and is such that the minimal gauge invariant fermion bilinear has axial charge $\pm 1$. This corresponds to the operator $n=1, m=0$. The axial symmetry, which is the shift symmetry in the bosonic description, is completely broken by the coupling $\frac{1}{2\pi} \phi F_{rt}$.}
The main difference from the Schwinger model is the position-dependent gauge coupling.

We proceed by integrating out $F_{rt}$, which leads to a potential for $\phi$
\begin{equation}\label{minnintout} S = \int_{r=0}^\infty dr\int dt\left(\frac{1}{8\pi } (\partial \phi)^2-\frac{e^2}{32\pi^3r^2}\min_{k\in \mathbb{Z}}(\phi+2\pi k)^2 \right)~. \end{equation} 
We see that the coupling to $A_r,A_t$ generates a centrifugal barrier with energy of order $e^2/r^2$. See also~\cite{Callan:1982au,Kazama:1983rt} and references therein. In addition a condensate is generated around the monopole $\langle \psi\tilde\psi \rangle \sim \frac{e}{r}$. The condensate breaks the axial symmetry, which is also explicitly broken to start with. (The computation of the prefactor in the condensate in the more familiar version of the Schwinger model, with a space-independent gauge coupling, is reviewed in~\cite{Smilga:1992hx}.)

We have only taken into account the effect of the zero modes of $A_r,A_t$ on the $S^2$. We need to justify why to this order in $e^2$ this suffices. Integrating out higher KK modes on the transverse sphere can produce quartic, sextic, and higher terms in the fermionic description. From the $SU(2)$ quantum numbers of the KK modes, we see that such terms can be obtained only at order $e^4$ and above. At order $e^2$ we can however induce a mass term for the $s$-wave fermions from integrating out KK modes, $\sim \frac{e^2}{r}\psi\tilde\psi$. This would be reflected in~\eqref{minnintout} as a cosine potential. To estimate the normalization we have to use the fermion condensate $\langle \psi\tilde\psi \rangle \sim \frac{e}{r}$, which would thus lead to a cosine potential $\sim \frac{e^3}{r^2}\cos(\phi)$. The sign of the cosine potential is not crucial at the moment since this is a small correction on top of the centrifugal barrier in~\eqref{minnintout}. 

Before we get into a more detailed discussion, it is intuitively clear that~\eqref{minnintout} implies a centrifugal repulsion of $\phi$ modes from the boundary at $r=0$. Therefore at the origin we will have to have $\phi\bigr|_{r=0}=0 \ {\rm mod}\  2\pi$. This has to be contrasted with the famous boundary condition~\eqref{eq:bar_box_n_1} 
$\psi\bigr|_{r=0} - e^{i\theta}\tilde \psi^\dagger=0$ which in the bosonic language would take the form $\phi=\theta \ {\rm mod} \  2\pi$. We see that the $\theta$ angle associated with monopole boundary conditions is not marginal when gauge field fluctuations are taken into account and only $\theta=0$ is possible. We will soon compute the scaling dimension of the associated defect operator.

On the branch $k=0$ of the potential in~\eqref{minnintout} we have the equation of motion   
\begin{equation}\label{Bessel}\left[\partial_t^2-\partial_r^2+\frac{e^2}{4\pi^2 r^2}\right]\phi =0\,.\end{equation}
Regardless of the frequency, the near-boundary behavior is given by $\phi\sim r^\lambda$ 
with $\lambda(\lambda-1) =\frac{e^2}{4\pi^2}$, which for small $e^2$ implies $\lambda = 1+\frac{e^2}{4\pi^2}+\mathcal{O}(e^4)$.
The coefficient of $r^\lambda$ is a defect operator of dimension $\Delta= 1+\frac{e^2}{4\pi^2}+\mathcal{O}(e^4)$ and hence it is an irrelevant defect operator. There is no exactly marginal $\theta$ angle on the monopole! 

The solution of~\eqref{Bessel} for nonzero frequency $w$ is given by 
\begin{equation}\label{Besssol}\phi=C\sqrt r J_{\rho}(rw)  ~,\end{equation}
with $\rho={1\over2}\sqrt{1+\frac{e^2}{\pi^2}}$, and $J_\rho$ is the Bessel function of the first kind. This leads to a phase shift for the $s$-wave 
$\delta_0 =-\frac{e^2}{8\pi} $. The phase shift is energy-independent due to the absence of mass scale in the system (the scale associated with the Landau pole is irrelevant at this order).

In summary, in QED$_4$ with one Dirac fermion and a minimal 't Hooft line at the origin $r=0$, we see that the gauge field fluctuations lead to a small but nonzero centrifugal barrier, the main effect of which is to remove the often-discussed marginal (at leading order in $e^2$) parameter $\theta$ from the boundary condition at the monopole. Another consequence of this effect is to introduce a small phase shift for scattering in the $s$-wave channel.

Consider now the case of QED$_4$ with one Dirac fermion but two units of monopole flux, i.e. $n=2$. 
We now have two incoming  fermions with charge 1 and two outgoing fermions of charge $-1$. We bosonize them as before and integrate out the gauge field to obtain the action
\begin{equation}\label{minnintoutn=2} S = \int_{r=0}^\infty dr\int dt\left(\frac{1}{8\pi } (\partial \phi_1)^2+\frac{1}{8\pi } (\partial \phi_2)^2-\frac{e^2}{32\pi^3r^2}\min_{k\in \mathbb{Z}}(\phi_1+\phi_2+2\pi k)^2 \right)~. \end{equation} 
The spectrum to this order in $e^2$ is one massive boson and one massless boson. If we substitute $\phi_1+\phi_2 = 0 \ {\rm mod} \ 2\pi$ in the action, we can see that the massless boson now has  $R^2=\frac{1}{2\pi}$, which is the self-dual radius. This will be important below.\footnote{This model is a variant, with a space-dependent gauge coupling, of the recently studied~\cite{Dempsey:2023gib} two-flavor QED$_2$. Another important difference is that we do not have a mass term, but instead, some quartics will be important. The effects of quartics in $2d$ were also considered in various recent papers~\cite{Cherman:2019hbq,Komargodski:2020mxz,Cherman:2022ecu,Delmastro:2022prj,Onder:2023tei}.}

Let us discuss the symmetries. To the order we are considering in $e^2$, since there are no quartics, the theory enjoys an accidental $SU(2)_L\times SU(2)_R$ which acts on the low energy massless boson at the self-dual point. 
However, the full QED$_4$ theory with two units of the elementary 't Hooft line only has a diagonal $SU(2)$ symmetry (corresponding to the rotations of the sphere) and an axial $\mathbb{Z}_4/\mathbb{Z}_2\sim \mathbb{Z}_2$ symmetry. The $\mathbb{Z}_2$ and the Cartan of $SU(2)$, which we denote by $U(1)_J$, act as
\begin{equation}
\mathbb{Z}_2: \ \phi_{1,2}\to\phi_{1,2}+\pi~,\quad U(1)_J: \ \phi_{1}\to\phi_{1}+C~,\phi_{2}\to\phi_{2}-C~.
\end{equation}
In terms of the original fermion variables we must allow quartic operators consistent with $SU(2)$ and $\mathbb{Z}_2$ axial symmetry. 
Indeed there are no bilinear operators due to the axial $\mathbb{Z}_2$ (this is unlike $n=1$). There are three quartics that are allowed by these symmetry considerations:
\begin{equation}\label{opers}\psi^i\psi^\dagger_i\tilde\psi^j\tilde\psi^\dagger_j~,
\qquad \epsilon_{ij}\epsilon_{kl}\psi^i\psi^j\tilde\psi^k\tilde\psi^l ~,\qquad  \epsilon_{ij}\epsilon^{kl}\psi^i\psi^\dagger_k\tilde\psi^j\tilde\psi^\dagger_l~, \end{equation}
where $i,j$ are the $SU(2)$ indices. 
Since the first and second operators preserve the left and right $SU(2)$ symmetries separately, they cannot lift the zero mode (or move the zero mode from the self-dual point). The third operator leads to a potential of the form $\cos(\phi_1-\phi_2)$ and simultaneously changes the boson radius, so that the diagonal $SU(2)$ symmetry is preserved. In terms of the compact boson effective theory at the self dual-point this acquires scaling dimension $\Delta=2$ and it is hence marginal, and the dynamics is similar to the BKT transition, which means that the final answer crucially depends on the sign of this quartic. 
For one sign we have a centrifugal barrier and a chiral condensate breaking $\mathbb{Z}_2$, (since this is a BKT transition, the associated scale is exponentially small) while for the other sign no centrifugal barrier or chiral condensate develop.
We leave the computation of this crucial sign to the future.

The fact that both choices of sign lead to a nontrivial phase (either a chiral condensate breaking the $\mathbb{Z}_2$ or a vanishing centrifugal barrier), is due to an 't Hooft anomaly in 2$d$.
Indeed this $\mathbb{Z}_2$ participates in a 2-group in QED$_4$ and hence has an 't Hooft anomaly in $2d$. This 't Hooft anomaly is elementary to detect directly in $2d$ (this is a repetition of the more general argument in footnote~\ref{DiscAno}): 
We can think of the axial $\mathbb{Z}_2$ as acting by a minus sign on the right-moving fermions only (this is equivalent to our original assignment by conjugating with a gauge transformation), then since we have 4 real right-moving fermions, the anomaly is 4 mod 8. The fact that this $\mathbb{Z}_2$ cannot be preserved in the presence of dynamical monopoles is also consistent with~\eqref{eq:QED_UV_symm}, where we have collected all the symmetries not participating in a 2-group.

For $N_f>1$ the presence of chiral isospin $SU(N_f)_L\times SU(N_f)_R$ symmetry in $2d$ implies an 't Hooft anomaly (descending, again, from the 2-group in $4d$) and hence protects some of the modes from developing a centrifugal barrier. Therefore, some fermions ought to be able to scatter from the actual monopole core. A simple scenario that is consistent with the anomalies is that the chiral $\mathbb{Z}_n$ is broken and we have in each of the vacua an $SU(N_f)_n$ WZW model of massless modes that can penetrate into the monopole core.

The problem of the fermion zero modes being lifted by gauge field fluctuations is similar in spirit to that of symmetric mass generation~\cite{EICHTEN1986179,Seiberg:1994bz,Seiberg:1994pq,Creutz:1996xc,Fidkowski:2009dba,Wang_2014,You_2018,Razamat:2020kyf,Tong:2021phe,Wang:2022ucy}, where fermions are gapped while preserving an anomaly-free chiral symmetry. In our situation there are some symmetries with 't Hooft anomalies and this is why the scenario that we mentioned, broken $\mathbb{Z}_n$ with an $SU(N_f)_n$ WZW model in each vacuum, is not trivially gapped (but still contains far fewer massless modes than to leading order in $e^2$). 

We end by briefly addressing the differences in the symmetry structure between the free fermion CFT and its gauged version, and how this slightly modifies the scattering story \eqref{eq:monopole_higher_n_fermonic_out} for $\operatorname{QED}(\ydiagram1,\ydiagram1)$, albeit only formally.

First, in a gauge theory, all operators must be gauge-invariant. Therefore the vertex operators $e^{iX_{i,\mu}}$, $e^{i\tilde{X}_{i,\mu}}$ used to construct ingoing and outgoing states in \eqref{eq:out-state_n} must be attached to Wilson lines going off to infinity, of charges $+1$ and $-1$ respectively, in order to make them gauge-invariant. Conservation of $U(1)_\text{gauge}$, which was regarded as part of the global symmetry \eqref{eq:QED_UV_symm} in the free fermion CFT, was justified because in the gauged theory, the total charge of the Wilson line going off to infinity cannot change. Similarly, the $T^{-2}$ line cannot terminate on a local operator in a gauge-invariant way, but must continue as a Wilson line of charge $+2$, with the twist operator $e^{-\frac{2}{nN_f}i\sum_{i,\mu}\tilde{X}_{i,\mu}}$ occurring on the junction. This results in the following slightly modified picture for the scattering, in which everything is attached to a Wilson line:

\begin{equation}
\begin{tikzpicture}[baseline=.9cm]

\draw[very thick] (-3,-0.6) -- (-3,2.3);
\foreach \x in {-4,...,14} \draw (-3-.2,.15*\x+.04) -- (.2-3-.2,.15*\x+.1+.04);
\foreach \x in {-5,...,13} \draw (-3-.1,.15*\x+.09+.075) -- (.2-3-.2,.15*\x+.1+.04+.075);

\filldraw (-1.2,1.6) circle (1pt);
\draw[->,>=stealth,thick] (-1.2,1.6) -- (-1.2+0.3,1.6+0.2);
\draw[dashed,postaction={decoration={markings,mark=at position 0.5 with {\arrow[scale=2]{>}}},decorate}] (-1.2,1.6) -- (2.4,1.6);
\node[anchor=south] at (-1.2+0.3/2,1.6+0.2/2) {$e^{i\tilde{X}_{i,\mu}}$};
\node[anchor=south] at (0.6,1.6) {$W^{-1}$};

\draw[decoration={snake,amplitude=.6mm},decorate] (-2.1,1.3) -- (-1.2, 1.3);
\path[postaction={decoration={markings,mark=at position 0.5 with {\arrow[scale=2]{>}}},decorate}] (-3,1.3) -- (-1.2,1.3);
\draw[decoration={snake,amplitude=.6mm},decorate] (-3,1.3) -- (-2.1, 1.3);
\filldraw (-1.2, 1.3) circle (1pt);
\draw[->,>=stealth,thick] (-1.2,1.3) -- (-1.2+0.3,1.3+0.2);
\draw[dashed,postaction={decoration={markings,mark=at position 0.5 with {\arrow[scale=2]{>}}},decorate}] (-1.2,1.3) -- (2.4,1.3);
\node[anchor=north] at (-2.1,1.3) {$T^{-2}$};
\node[anchor=north] at (0.6,1.3) {$W^2$};

\filldraw (-1.2,-0.2) circle (1pt);
\draw[->,>=stealth,thick] (-1.2,-0.2) -- (-1.2-0.3,-0.2+0.2);
\draw[dashed,postaction={decoration={markings,mark=at position 0.5 with {\arrow[scale=2]{>}}},decorate}] (-1.2,-0.2) -- (2.4,-0.2);
\node[anchor=south] at (-1.2+0.3/2,-0.2+0.2/2) {$e^{iX_{i,\mu}}$};
\node[anchor=south] at (0.6,-0.2) {$W$};

\end{tikzpicture}
\end{equation}

Second, in a gauge theory, gauge transformations act trivially. This leads to a slight puzzle regarding the $T^{-2}$ line attached to the outgoing radiation. Recall that the $T^{-2}$ line is defined as an element of \eqref{eq:QED_IR_symm_New} by a combination of an axial transformation $-2 \in \Z_{2nN_f}$ and a gauge transformation $e^{2\pi i / nN_f} \in U(1)_\text{gauge}$. The gauge transformation was chosen to undo the effect of the axial transformation on the left-moving fermions, ensuring that the combination acts purely on the right-moving fermions,
\be
\psi_{i,\mu} \rightarrow \psi_{i,\mu}\,,
\qquad
\tilde{\psi}_{i,\mu} \rightarrow e^{-4\pi i / nN_f} \tilde{\psi}_{i,\mu}\,.
\ee
This was crucial to ensure that $T^{-2}$ created a right-moving (outgoing) excitation, as it should.

If instead we had chosen the opposite gauge transformation, then the combination would act purely on the left-movers,
\be
\psi_{i,\mu} \rightarrow e^{-4\pi i / nN_f} \psi_{i,\mu}\,,
\qquad
\tilde{\psi}_{i,\mu} \rightarrow \tilde{\psi}_{i,\mu}\,.
\ee
In the gauged theory, these choices are gauge equivalent, so should be indistinguishable. But at least in the free fermion CFT, the former creates an outgoing excitation, while the second creates an ingoing excitation, and these choices are certainly physically distinguishable. So how can a gauge choice make a physical difference? The resolution is that when the above two transformations are placed on an open line, ending at $r = r_0$, the transformation from one to the other induces a change in the dynamical gauge field
\be
A_r \rightarrow A_r - \tfrac{4\pi}{n N_f} \delta(r - r_0)\
\ee
so they are not gauge equivalent. Thus the $T^{-2}$ line in the gauged theory can end on either a left-moving or right-moving operator, and we should make the choice that results in a right-moving operator. Furthermore, beyond leading order in $e^2$, the $e^2/r^2$ potential \eqref{minnintout} mixes left-movers and right-movers, so there is no sharp distinction between them anyway.

\section{Non-invertible Surface in $\boldsymbol{4d}$}\label{sec:uplift_4d}

A result from section~\ref{sec:non-chiral_4d} is that if we throw $\psi$ into the boundary, what comes out is $\tilde\psi$, attached to a topological line. Lifting this picture to the $4d$ theory, an in-coming $j_0$-wave becomes an outgoing $j_0$-wave that lives at the end of a $3d$ topological surface. What does this surface correspond to? The $2d$ defect acts on the right-moving fermions as a discrete rotation $e^{-4\pi i/nN_f}$. Composing this with a gauge transformation, we can also write this as a rotation on both left-movers and right-movers simultaneously, which in $4d$ amounts to a chiral rotation
\begin{equation}
\label{eq:lineaction4d}
U(1)_A\colon \Psi_i\mapsto e^{\frac{2\pi i}{nN_f}\gamma^5}\Psi_i\,.
\end{equation}
But note that this is not a standard symmetry of the $4d$ theory. Instead, the invertible $0$-form chiral symmetry is just the $\mathbb Z_{2N_f}\subset U(1)_A$ subgroup (which does not have ABJ anomalies). The final conclusion is, then as follows: for $n>2$ the outgoing state is attached to the non-invertible symmetry defect $\cD_{p/N}$ introduced in~\cite{Choi:2022jqy,Cordova:2022ieu}, which acts on local operators in the same way the naive $U(1)_A$ would, but also non-trivially (and non-invertibly) on 't Hooft lines.

Let us recall how the defect $\mathcal D_{p/N}$ is constructed, following~\cite{Choi:2022jqy,Cordova:2022ieu}.\footnote{See also~\cite{Karasik:2022kkq,GarciaEtxebarria:2022jky} for an alternative perspective.} As is well-known, the axial rotation in QED 
\be 
\label{eq:axialtrans}
U(1)_A\colon \Psi_i \mapsto e^{i \alpha \gamma_5/2} \Psi_i\,, \quad \alpha \simeq \alpha+2\pi\,,
\ee 
suffers an ABJ anomaly~\cite{PhysRev.177.2426,Bell:1969ts}
\be 
\mathrm d \star j^A=\frac{N_f}{4\pi^2} F \wedge F\,,
\ee 
where the axial current is
\be 
j^A_\mu= \sum_i \bar \Psi_i \gamma_5 \gamma_\mu \Psi_i\,.
\ee 
The flux $F=\mathrm dA$ is normalized as $\int_\Sigma F \in 2\pi \Z$ for a closed surface $\Sigma$, e.g.,~a sphere centered at the monopole will measure the magnetic charge $\int_{S^2} F=2\pi n$.
Due to the anomaly, the usual axial operator $e^{\frac{i\alpha}{2} \oint \star j^A}$ is not topological in the quantum theory. One could try to define a topological symmetry operator as
\be 
\label{eq:non-gaugeinvop}
U_\alpha=\exp \lbb \frac{i\alpha}{2} \oint \star j^A-\frac{N_f}{4\pi^2} A \wedge \mathrm dA \rbb\,,
\ee 
but this operator is only gauge-invariant when $\alpha$ is a multiple of $2\pi /N_f$, since otherwise the Chern-Simons term is improperly quantized. Therefore $U_{2\pi/N_f}$ only generates an invertible $\Z_{2N_f} \subset U(1)_A$ symmetry. 

There is in fact an additional discrete, non-invertible component leftover from the axial transformation, generated by topological defects $\cD_{p/N}$ labeled by a rational number $p/N \in \Q/\Z$. 
The topological operator $\cD_{p/N}$ (also denoted as $\mathcal D_k$, with $p/N\equiv N_f/k$) is composed of an axial rotation by $\alpha=\frac{2\pi}{N_f}\times \frac{p}{N}$ and a fractional quantum Hall state of filling fraction $\nu_\text{fill}=p/N$ 
\be 
\label{eq:noninvop}
\cD_{p/N}=\exp \lbb \oint \frac{i \pi }{N_f} \times \frac{p}{N}\star j^A+\cA^{N,p}[dA/N] \rbb\,.
\ee
Here $\cA^{N,p}$ is the minimal TQFT \cite{Hsin:2018vcg} with $1$-form symmetry $\Z^{(1)}_N$ and 't Hooft anomaly labeled by $p \simeq p+N$. The notation $\cA^{N,p}[B]$ refers to coupling to a background $2$-form gauge field $B$ for the $1$-form symmetry. In~\eqref{eq:noninvop} we couple to the bulk gauge field $B=dA/N$. 
If $\gcd(p,N)=1$ the theory $\cA^{N,p}$ forms a consistent $3d$ TQFT on its own. 

The action \eqref{eq:lineaction4d} determines that the outgoing radiation lives in the sector twisted by the $\cD_{p/N}$ defect\footnote{A caveat is that the non-invertible defect is not uniquely determined by its action on the fermions. In principle $\cA^{N,p}[B]$ could be replaced by any TQFT $\cT[B]$ which has a $\Z_N$ $1$-form symmetry with anomaly $p$. It was shown in \cite{Hsin:2018vcg} that such a TQFT factorizes as $\cT[B]=\cA^{N,p}[B] \times \cT'$, where $\cT'$ is a decoupled TQFT whose lines are neutral under the $1$-form symmetry of $\cA^{N,p}$. Whether or not such a decoupled sector $\cT'$ lives on the outgoing defect does not appear to be determinable by this experiment alone, due to the trivial topology of the surface.} with 
\be 
p/N=2/n \  \text{mod} \ 1\,.
\ee
The minimal abelian TQFT $\cA^{N,p}$ is only unitary for $\gcd(N,p)=1$, so 
\be 
\label{eq:Npidentification}
N,p=\begin{cases}
	\frac{n}{2}, 1 & n \ \text{even}\\
	n, 2 & n \ \text{odd}
\end{cases}
\ee 
The symmetry defect is non-invertible for all $n>2$.

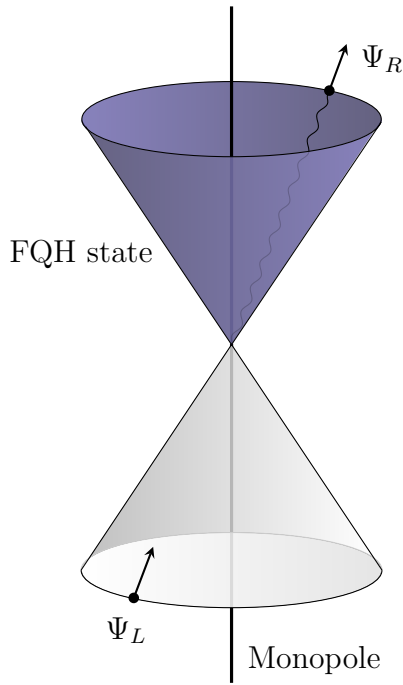
\begin{figure}[h!]
\centering
\begin{tikzpicture}
\draw[very thick] (0,-7.5) -- (0,-2.99);
\fill[left color={rgb,255:red,75;green,69;blue,129},right color={rgb,255:red,144;green,133;blue,250},opacity=.5] (2,0) -- (0,-3) -- (-2,0) arc (180:360:2cm and 0.5cm);
\fill[left color=gray!50!white,right color=white,opacity=1] (2,-6) -- (0,-3) -- (-2,-6) arc (180:360:2cm and 0.5cm);
\draw[very thick,color={rgb,255:red,75;green,69;blue,129},opacity=.4] (0,-3) -- (0,-.5);
\draw[very thick,color=black,opacity=.1] (0,-3) -- (0,-6.5);
\fill[right color={rgb,255:red,75;green,69;blue,129},left color={rgb,255:red,144;green,133;blue,250},opacity=.5] (0,0) circle (1.99cm and 0.5cm);
\draw[very thick] (0,-.5) -- (0,1.5);
\draw (-1.94,-.1) -- (0,-3) -- (1.99,-.04);
\fill[right color=gray!20!white,left color=white] (0,-6) circle (1.99cm and 0.5cm);
\draw[very thick,color=black,opacity=.1] (0,-3) -- (0,-6.5);
\draw (0,0) circle (1.99cm and 0.5cm);
\draw (-2,-6) arc (180:360:2cm and 0.5cm);
\draw[black!20!white] (-2,-6) arc (180:0:2cm and 0.5cm);
\draw (-2,-6) -- (0,-3) -- (2,-6);
\fill (-1.3,-6.37) circle (2pt); \draw[thick,->,>=stealth] (-1.3,-6.37) -- (-1.3+.258,-6.37+.68);
\fill (1.3,.38) circle (2pt); \draw[thick,->,>=stealth] (1.3,.38) -- (1.53,1);
\begin{scope}
\clip (2,0) -- (0,-3) -- (-2,0) arc (180:360:2cm and 0.5cm);
\draw[decorate, decoration={snake,amplitude=.4mm},opacity=.2] (0,-3) -- (1.3,.38);
\end{scope}
\begin{scope}
\clip (0,0) circle (1.99cm and 0.5cm);
\draw[decorate, decoration={snake,amplitude=.4mm}] (0,-3) -- (1.3,.38);
\end{scope}
\node at (1.1,-7.2) {Monopole};
\node at (-2,-1.8) {FQH state};
\node at (-1.4,-6.8) {$\Psi_L$};
\node at (2,.8) {$\Psi_R$};
\end{tikzpicture}
\label{fig:anyon_FHS}
\caption{Spacetime diagram for the scattering process of the lowest-lying partial wave mode with one angular dimension suppressed. The left-handed component $\Psi_L$ is a local in-falling excitation, whereas the outgoing right-handed component $\Psi_R$ is trailed by a surface with a Wilson line excitation.}
\end{figure}

To summarize the discussion so far, we have seen that, in QED$_4$, a scattering process involving a heavy monopole and a low-energy massless fermion gives rise to an out-state consisting of a spherically symmetric wave of fermions that lives on the boundary of the $3d$ topological defect $\mathcal D_{2/n}$. This is not the end of the story however, since we have yet to offer a $4d$ picture for the electric vacuum charge of the outgoing surface: at face value, the incoming fermion carries charge $+1$ while the outgoing fermion carries charge $-1$, and it is not clear how the defect accounts for the missing charge. The upshot will be that the outgoing fermion is attached to an anyon of $\mathcal A^{N,p}$ which adds the two units of missing electric charge; see figure~\ref{fig:anyon_FHS}.

In $3d$, any abelian TQFT (including in particular our Hall state $\mathcal A^{N,p}$) admits a realization in terms of a Chern-Simons theory over a torus,
\begin{equation}\label{eq:general_Hall}
\mathcal L_{\mathcal A^{N,p}}(a,A)=\frac{1}{4\pi} a^t K \mathrm da+\frac{1}{2\pi} a^t v\ \mathrm d A\,,
\end{equation}
where $a$ is a tuple of $3d$ $U(1)$ gauge fields defined on the worldvolume of $\mathcal D$, $K$ is an integral symmetric matrix, $A$ is the $4d$ electromagnetic field, and $v$ is an integral vector that dictates how $a$ couples to $A$. The filling fraction for a given $K,v$ can be written as $p/N=v^tK^{-1}v$. For example, a Lagrangian for $\mathcal A^{N,1}$ can be written as~\eqref{eq:general_Hall} with $1\times1$ matrix $K=(N)$ and $v=1$, cf.
\begin{equation}\label{eq:D1/p}
\mathcal L_{\mathcal A^{N,1}}(a,A)= \frac{N}{4\pi}a\, \mathrm da+\frac{1}{2\pi}a\, \mathrm dA\,.
\end{equation}
See appendix~\ref{app:hall_TQFT} for the form of $K,v$ for arbitrary filling fraction $p/N$.

The source of electric charge on the surface becomes entirely clear when we consider the partition function of the TQFT on a sphere with $n$ units of magnetic flux. 
Using the expression for the abelian TQFT~\eqref{eq:general_Hall}, the partition function becomes
\begin{equation}
Z_{\mathcal A^{N,p}}[n]=\int D a \ \exp \biggl[i\int \frac{1}{4\pi}a^t K \mathrm da+ i n \int_{\gamma} a^tv \biggr]\,,
\end{equation}
where $\gamma$ is the worldline of the fermion in the $(r,t)$ directions. This has a simple interpretation as the expectation value of the anyon $W^n$ smeared over the $S^2$, where $W=e^{iv^t\int a}$ is the generating line of the fusion algebra $\mathbb Z_N^{(1)}$. This Wilson line carries electric charge $nv^tK^{-1}v\equiv np/N$, which accounts for the two units of missing electric charge since for our defect $np/N=2$. 
As in the $2d$ case, this means that in order to maintain gauge invariance, $W^n$ must continue as a charge-$2$ electromagnetic Wilson line $e^{2i\int A}$ after exiting the TQFT, stretching to infinity.

The exact details of the excitation created at the boundary of $\mathcal{D}_{2/n}$ will depend on the boundary state chosen for the TQFT, which so far we have left implicit. To give a more explicit description, let us take $\mathcal{D}_{2/n}$ to be supported on a 3-ball ``pancake'' of radius $r_0$. Recall that $\mathcal{D}_{2/n}$ acts invertibly on the fermions as \eqref{eq:axialtrans}, and by composing with a suitable gauge transformation, we can take it to act purely on right-handed fermions, as
\be\label{eq:d2ntransformpsix}
\mathcal{D}_{2/n}:\quad \psi_{R,i}(\mathbf{x}) \rightarrow \psi_{R,i}(\mathbf{x}) \begin{cases}e^{-4\pi i/nN_f} & |\mathbf{x}| < r_0 \\ 1 & |\mathbf{x}| > r_0\end{cases}\,.
\ee
Although this gauge choice makes no difference to $\mathcal{D}_{2/n}$ on a closed manifold, as in section~\ref{sec:gauged}, it does affect the choice of twist operators we place at the boundary. For example, in the lowest partial wave, it ensures that the twist operators are purely out-moving. The transformation \eqref{eq:d2ntransformpsix} also acts on the higher partial waves $j > j_0$, and therefore creates both in-moving and out-moving twist operators there. They have a total electric charge of zero since the higher partial waves are non-chiral, and furthermore, are suppressed from exciting propagating modes near the monopole due to the centrifugal barrier \eqref{eq:centrifuge}.

In conclusion, we find that the outgoing fermion lives at the end of a Wilson line that emanates from the monopole. The local fermion that the surface ends on carries charge $-1$. The line adds two units of electromagnetic charge, making the total charge of the out-state $+1$, which nicely matches the charge of the ingoing fermion. This way, we see that the scattering process conserves electric charge, as it should. We stress that the Wilson line is transparent in $\mathcal A^{N,p}$, and therefore it does not lead to anyonic statistics for the $4d$ fermions (it also has integral spin and therefore does not change the fermion number of the out-state).

\begin{center}***\end{center}

Having presented a comprehensive proposal for the outgoing state of the fermion-monopole scattering, for which most our evidence came from a 2$d$ reduction, let us briefly consider the argument from a purely 4$d$ physics perspective. 
The key idea is that the outgoing fermions live at the boundary of a topological defect that carries electric charge, and which has a topological junction with the 't Hooft line so it must generate some subset of the symmetry preserved by the monopole. 
In QED$_4$ there is only one reasonable candidate for this defect, namely $\mathcal D_{p/N}$ for some $p/N$. And as we have just shown, such a defect surrounding the monopole necessarily carries an electromagnetic Wilson line of electric charge $np/N$, and we know that its action commutes with $G$, which are the qualities we are looking for. 

Above, we identified $p/N$ by matching with the action of the $2d$ line, but in fact we might have determined it by noting that the $\cD_{p/N}$ surface attached to the monopole has a $1$-form symmetry anomaly~\cite{Hsin:2018vcg}
\be
- \frac{ip}{4\pi N} \int F \wedge F\,,
\ee
which integrates to 
\be
- \frac{ip n}{4\pi N} \int F\,,
\ee
on $S^2$. (This is simply a reformulation of footnote~\ref{SymmFoot}.)  As pointed out in~\cite{Delmastro:2022pfo}, the presence of this anomaly implies a vanishing partition function for the $1d$ theory obtained by compactifying $\cA^{N,p}[dA/N]$ on $S^2$, unless $pn=\ell N$ for some $\ell\in\Z$\footnote{This is because a background gauge transformation for the $1$-form symmetry induces a rotation of the partition function by a phase, which is only consistent if the partition function vanishes (in general this is true of any anomalous global $d$-form symmetry in $d$ dimensions).} which is the electric charge of the Wilson line in the TQFT. To determine $\ell$, we simply note that the electric charge discrepancy between $\psi$ and $\tilde \psi$ is $+2$, and an outgoing surface $\cD_{2/n}$ supports a TQFT with a Wilson line of electric charge $+2$. This simple argument directly identifies $\mathcal D_{2/n}$ as the most natural twist defect that is attached to the outgoing radiation, without having to go through the full details of the $2d$ reduction.

\section{Chiral Theories and Other Future Directions}\label{sec:chiral}

We close with some comments on the application of the ideas above in chiral theories. Consider a $4d$ gauge theory with gauge group $U(1)^{N_c}$ and a collection of left-handed Weyl fermions $\chi_\alpha$ with charges $Q_{\alpha a}$, where $\alpha=1,2,\dots, N_f$ and $a=1,2,\dots, N_c$. The Lagrangian is
\begin{equation}
\mathcal L=-\frac12\sum_{a,b=1}^{N_c} e^{-2}_{ab}F_a\wedge\star F_b+\sum_{\alpha=1}^{N_f} \chi^\dagger_\alpha \sigma^\mu\bigl(i\partial_\mu+
\sum_{a=1}^{N_c}Q_{\alpha a}A_{\mu,a}\bigr)\chi_\alpha\,,
\end{equation}
with $e_{ab}$ the coupling constants. Gauge transformations act as $A_a\mapsto A_a+\mathrm d \Gamma_a$ and $\chi_\alpha\mapsto e^{i\sum_{a=1}^{N_c}Q_{\alpha a}\Gamma_a}\chi_\alpha$, with $\Gamma_a$ a compact scalar. There are no theta terms since the fermions are massless (and we assume $\operatorname{rank}(Q)=N_c$ so there are no decoupled photons). Vector-like theories are recovered by taking the Weyl fermions to come in pairs of opposite charges $Q_{\alpha,a}=-Q_{\alpha',a}$ and packaging them into Dirac fermions.

The gauge anomaly cancellation conditions are $\mathcal A_{abc}=0$ and $\mathcal A_a=0$, where
\begin{equation}
\begin{alignedat}{2}
&\tr[U(1)_aU(1)_bU(1)_c]\colon\ \mathcal A_{abc}&&=\sum_\alpha Q_{\alpha a}Q_{\alpha b} Q_{\alpha c}\\
&\tr[U(1)_a]\colon\hspace{2.3cm} \mathcal A_a&&=\sum_\alpha Q_{\alpha a}
\end{alignedat}
\end{equation}
The condition $\mathcal A_a=0$ guarantees that the theory can be placed on a curved space consistently. Alternatively, it guarantees that there is no $2$-group between the Poincar\'e symmetry and the $1$-form magnetic symmetry.
 
We insert a monopole with magnetic charges $m_a\in\mathbb Z$, defined by removing a point from spacetime and imposing boundary conditions on the gauge field, schematically of the form
\begin{equation}
A_a\sim \frac12m_a(1-\cos\theta)\mathrm d\phi\,.
\end{equation}
If we repeat the same analysis as in section~\ref{sec:non-chiral_4d}, we find that each $4d$ Weyl fermion gives rise, in its lowest angular momentum mode, to a $2d$ fermion with the same charge $Q_{\alpha a}$ under $U(1)^{N_c}$, chirality $\gamma_\alpha^\star=\sign (Qm)_\alpha$ and $SU(2)_J$ representation of dimension $|(Qm)_\alpha|$ (i.e., with spin $j_0=\frac{|(Qm)_\alpha|-1}{2}$).

Let us check that the $2d$ 't Hooft anomalies vanish for all the symmetries preserved by the monopole; this will allow a boundary condition that is compatible with such symmetries. For example, the quadratic anomaly for the gauge group $U(1)^{N_c}$ is\footnote{Here $\mu$ denotes the $SU(2)_J$ weight and it takes values $\mu=-j_0,-j_0+1,\dots,j_0$, with $j_0=\frac{|(Qm)_\alpha|-1}{2}$.}
\begin{equation}
\begin{aligned}
\tr[U(1)_aU(1)_b]\colon \sum_{\alpha,\mu}Q_{\alpha a}Q_{\alpha b}\gamma_\alpha^\star&=\sum_\alpha |(Qm)_\alpha|Q_{\alpha a}Q_{\alpha b}\sign (Qm)_\alpha\\
&=\sum_{\alpha,c} Q_{\alpha c} m_c Q_{\alpha a}Q_{\alpha b}\\
&=\sum_{c}\mathcal A_{abc}m_c\,,
\end{aligned}
\end{equation}
which indeed vanishes if $\mathcal A_{abc}=0$. The $2d$ gravitational anomaly is 
\begin{equation}
\begin{aligned}
c_L-c_R\colon\sum_{\alpha,\mu} \gamma_\alpha^\star&=\sum_\alpha |(Qm)_\alpha|\sign (Qm)_\alpha\\
&=\sum_{\alpha,a} Q_{\alpha a}m_a\\
&=\sum_{a} \mathcal A_a m_a\,,
\end{aligned}
\end{equation}
which again vanishes if $\mathcal A_a=0$.

More generally, a $4d$ mixed anomaly of the form $G$-$G$-$U(1)_a$ for some flavor group $G$ induces, after the reduction to the lowest angular momentum mode, a $G$-$G$ 't Hooft anomaly in $2d$. If this anomaly is non-zero, the symmetry $G$ will not be preserved by the boundary. (However, a subgroup of $G$ may be preserved.) We therefore learn that the scattering experiment will preserve all symmetries that do not participate in a $2$-group. This is as expected, since the monopole acts as a source for the magnetic $1$-form symmetry participating in the $2$-group.

Finally, let us show that $SU(2)_J$ is also anomaly-free, which is a necessary condition for the existence of spherically-symmetric monopoles. Recall that a $2d$ $G$-$G$ anomaly is measured by the Dynkin index of the representation under which the fermions transform; for the $d$-dimensional representation of $SU(2)$, $T(d)=\frac{1}{12}d(d^2-1)$. Therefore, the anomaly of $SU(2)_J$ is
\begin{equation}
\begin{aligned}
\sum_\alpha T(|(Qm)_{\alpha}|)\gamma_\alpha^\star&=\frac{1}{12}\sum_\alpha (Qm)_{\alpha}((Qm)_{\alpha}^2-1)\\
&=\frac{1}{12}\sum_{a,b,c} \mathcal A_{abc}m_am_bm_c-\frac{1}{12}\sum_a \mathcal A_a m_a\,,
\end{aligned}
\end{equation}
which again vanishes if the linear and cubic gauge anomalies vanish.

The rules of the game are, then, as follows. Given an arbitrary $4d$ abelian gauge theory, a scattering process in the presence of a monopole of magnetic charge $m_a\in\mathbb Z$ reduces, at low energies, to an effective $2d$ boundary problem. Each $4d$ Weyl fermion of charge $Q_{\alpha a}$ leads to a $2d$ free fermion with the same gauge charge, chirality $\gamma_\alpha^\star=\sign (Qm)_\alpha$, and $SU(2)_J$ spin $j_0=\frac{|(Qm)_\alpha|-1}{2}$. The boundary condition at the monopole conserves all the symmetries of the original $4d$ theory that do not participate in a $2$-group. Then, a typical scattering process consists of sending a left-moving fermion towards the boundary, which will become a right-moving excitation with the same quantum numbers; such an outgoing state is, typically, a twist operator that lives at the end of some topological line. In $4d$, this is interpreted as outgoing radiation that is attached to a $3d$ topological surface. This surface can be invertible or non-invertible, depending on the details of the theory.

This overall picture is quite similar to the non-chiral example discussed in the previous sections. That being said, chiral theories do give rise to some special features. A simple example can be illustrated via the gauge theory consisting of a single $U(1)$ gauge field and five fermions with charges $-1,-5,7,8,-9$. As is easily checked, this theory has no cubic or linear gauge anomalies. The $2d$ reduction, in the presence of a minimally charged monopole, consists of three right-moving fermions with charges $-1,-5,-9$, and two left-moving fermions with charges $7,8$. These fermions transform in the $SU(2)_J$ representations $\boldsymbol1,\boldsymbol5,\boldsymbol9$ and $\boldsymbol7,\boldsymbol8$, respectively. Consider now what happens if we send the left-mover with charge $8$ towards the boundary; this fermion transforms in the $\boldsymbol 8$ of $SU(2)_J$, which is even-dimensional. On the other hand, all the right-movers transform in odd-dimensional representations of $SU(2)_J$. Therefore, conservation of $SU(2)_J$ requires the topological line itself to carry $SU(2)_J$ charge. This is only possible if the line either does not commute with $SU(2)_J$, or is non-invertible, or it carries fermionic zero-modes (recall that invertible lines without zero-modes never add charge under semi-simple groups, cf.~footnote~\ref{fn:semi-simple}). 
In any case, we see that the out-states in chiral theories can be more involved than those of vector-like theories.
This feature is actually of some phenomenological relevance. It turns out that certain monopoles of $SU(5)$ GUT theories lead to a spectrum of $2d$ fermions that cannot conserve the rotation symmetry unless we attach a topological line that is as in the chiral example above. We leave the details to future work.

\subsection{Additional Open Questions}

Finally, we highlight a few open questions.

\begin{enumerate}

\item The approach we took throughout this paper is that the monopole is infinitely heavy and has an internal gap (for instance a gap to a dyon state or other possible internal states). It would be interesting to reconsider the problem when this is not true, for instance, if there are normalizable zero modes on the 't Hooft line. This corresponds in the $2d$ reduction to a non-simple boundary condition!

\item We have assumed throughout the paper that monopoles are represented by simple boundary conditions. This forced the breaking of $\mathbb{Z}_{nN_f}$ to $\mathbb{Z}_{nN_f/\gcd(2,nN_f)}$ by the boundary condition. Instead, a non-simple boundary condition can preserve the full axial $\mathbb{Z}_{nN_f}$. It would be interesting to understand how these different scenarios occur in various UV complete theories. A simple example of a theory that should lead to non-simple monopole boundary conditions is $SO(3)$ $\mathcal{N}=1$ supersymmetric Yang-Mills theory (see~\cite{Kaidi:2021xfk,Argurio:2023lwl} for nice discussions of the axial symmetry of this theory).

\item Is there a similar story for scattering in the presence of axions~\cite{Choi:2022fgx}? What about in the $\mathbb{CP}^1$ NLSM in $3+1$ dimensions~\cite{Chen:2022cyw}? The $\mathbb{CP}^1$ NLSM in $3+1$ dimensions has a $1$-form symmetry and some non-invertible surfaces, therefore, it would be interesting to see if they may play a similar role.
\item Our discussion did not determine the precise topological surface that comes out from the scattering gedanken experiment (only the $1$-form symmetry anomaly of the FQH state). Can it be determined in some UV completions of QED$_4$? 

\item Related to the previous point, how do we excite the non-transparent anyons of the Hall state? What is their physical significance?
\item Another interesting generalization is to use fermions of higher electric charge. A new feature of such theories is that there is an electric $\mathbb Z^{(1)}_q$, that mixes with the rest of the symmetries in a non-trivial way. The reduced $2d$ theory now is a variant of the Schwinger model with higher charges, which has its own new features~\cite{Komargodski:2017dmc,Honda:2022edn,Cherman:2022ecu}. 

\item We have left the question of determining which scattering states  have no centrifugal barrier to the future. It would be very nice to understand that along with the condensates that are turned on in the monopole vicinity. In the language of defect conformal field theory this is equivalent to determining the space of marginal defect operators and the one point functions of bulk operators. The modes with no centrifugal barrier that we discussed in the text for $N_f>1$ are tilt operators in the language of defect conformal field theory. 

\end{enumerate}

\section*{Acknowledgement}

We thank T.D.~Brennan, Y.~Choi, G.~Cuomo, N.~Dorey, K.~Ohmori, N.~Seiberg, A.~Sever, S-H.~Shao, M.~Watanabe, S.~Yankielowicz and Y.~Zheng for useful discussions.
ZK is supported in part by the Simons Foundation grant 488657 (Simons Collaboration on the Non-Perturbative Bootstrap) and the BSF grant no. 2018204. DT is supported by the STFC grant ST/L000385/1, the EPSRC grant EP/V047655/1 ``Chiral Gauge Theories: From Strong Coupling to the Standard Model", and a Simons Investigator Award. PBS is supported by WPI Initiative, MEXT, Japan at Kavli IPMU, the University of Tokyo.

\appendix

\section{Some technical computations in $\boldsymbol{2d}$ free fermion CFT}\label{app:scaling_ferm}

In this appendix we collect some computations regarding free fermion CFTs.

\paragraph{Boundary States for the 3450 Model.}

In this subsection we sketch the classification of all conformal boundaries for the 3450 model, i.e., those preserving only Virasoro and $U(1)$ symmetry. The basic idea is to perform the coset construction by the $U(1)$, yielding a Virasoro algebra at $c=1$ that must also be preserved by the boundary. Since the complete classification of boundary states at $c=1$ is known \cite{Friedan:1999,Janik:2001hb}, this leads to the classification for the 3450 model. The results are as follows: in addition to the $U(1) \times U(1)'$- and $U(1) \times U(1)''$-preserving states described in the main text, there exists a small family of pathological states, the so-called Janik states~\cite{Janik:2001hb}, which have continuous spectrum in the open sector and are thus usually ruled out. Ignoring these states, this justifies our assumption that an extra $\Z_5$ is preserved by the boundary.

We start by decomposing the partition functions of the left- and right-movers into characters under the relevant chiral algebras. In our case, these are a $U(1)_5$ current algebra from the preserved $U(1)$ symmetry, and a commuting $\text{Vir}_{c=1}$ algebra from the coset construction. These have characters
\be
\chi_n^{U(1)_5}(\tau, a) = e^{ina} \frac{q^{n^2/50}}{\eta(\tau)}\,, \qquad \chi_h^{\text{Vir}_{c=1}}(\tau) = \frac{q^h}{\eta(\tau)}
\ee
where $a \sim a + 2\pi$ is a $U(1)$ fugacity. In terms of these, the partition functions are
\begin{equation}
\begin{aligned}
\mathcal{Z}_L(\tau, a) &= \sum_{n_1, n_2 \in \Z} \chi_{3n_1 + 4n_2}^{U(1)_5}(\tau, a) \chi_{(4n_1 - 3n_2)^2 / 50}^{\text{Vir}_{c=1}}(\tau)\,,\\
\mathcal{Z}_R(\tau, a) &= \sum_{\tilde{n}_1, \tilde{n}_2 \in \Z} \chi_{5\tilde{n}_1}^{U(1)_{5}}(\tau, a) \chi_{\tilde{n}_2^2/2}^{\text{Vir}_{c=1}}(\tau)\,.
\end{aligned}
\end{equation}
All the characters appearing in $\mathcal{Z}_{L,R}$ are distinct irreducibles, with the sole exception of $\chi_0^{\text{Vir}_{c=1}}(\tau)$, which splits into irreducibles as
\be
\chi_0^{\text{Vir}_{c=1}}(\tau) = \sum_{J=0}^{\infty }\frac{q^{J^{2}} - q^{(J+1)^2}}{\eta(\tau)}\,.
\ee
We wish to find all irreducibles $\mathcal{M}_{L} \in \mathcal{Z}_L$ and $\mathcal{M}_{R} \in \mathcal{Z}_R$ such that $\mathcal{M}_L \cong \mathcal{M}_R$. Ignoring the reducibility of $\chi_0^{\text{Vir}_{c=1}}$ at first, these are precisely the solutions to
\begin{equation}
\begin{aligned}
3n_1 + 4n_2 &= 5\tilde{n}_1\,,\\
4n_1 - 3n_2 &= \pm 5\tilde{n}_2\,.
\end{aligned}
\label{eq:nslns}
\end{equation}
Each solution gives one Ishibashi state preserving $U(1)$ and Virasoro. (This is essentially the rigorous version of the argument given in section~\ref{sec:u1z5andbeyond} based on conservation of scaling dimension.) However, when $\tilde{n}_2 = 0$, reducibility of $\chi_0^{\text{Vir}_{c=1}}$ implies that we actually obtain an infinite set of Ishibashi states labeled by $J \geq 0$.

To write down the Ishibashi states, let us split the solutions to \eqref{eq:nslns} into three subsets: $S_0$ and $S_\pm$, with $S_0$ consisting of the solutions with $\tilde{n}_2 = 0$, and $S_\pm$ consisting of the solutions with $\tilde{n}_2 \neq 0$ and the $\pm$ choice of sign in \eqref{eq:nslns}. Then the Ishibashi states are
\be
\Vert (n_1, n_2, \tilde{n}_1, \tilde{n}_2) \in S_\pm \rangle\!\rangle\,, \qquad \Vert (n_1, n_2, \tilde{n}_1, 0) \in S_0; J \geq 0 \rangle\!\rangle\,.
\label{eq:allishis}
\ee
The boundary states preserving an additional $U(1)$ symmetry can be written in terms of these states as
\begin{equation}
\begin{aligned}
\frac{1}{\sqrt{5}} |\cB_\pm(\theta_1, \theta_2)\rangle &= \sum_{(n_1, n_2, \tilde{n}_1, \tilde{n}_2) \in S_\pm} e^{i (\tilde{n}_1 \theta_1 + \tilde{n}_2 \theta_2)} \Vert (n_1, n_2, \tilde{n}_1, \tilde{n}_2) \rangle\!\rangle\ \\
&+ \sum_{(n_1, n_2, \tilde{n}_1, 0) \in S_0} \sum_{J=0}^\infty e^{i \tilde{n}_1 \theta_1} (\pm 1)^J \Vert (n_1, n_2, \tilde{n}_1, 0); J \rangle\!\rangle \,,
\end{aligned}
\end{equation}
where the choice of sign corresponds to whether $U(1)'$ or $U(1)''$ is the additional preserved symmetry. But these boundary states do not quite saturate the available Ishibashi states \eqref{eq:allishis}; the remaining states take the form of Janik-like states
\be
\frac{1}{\sqrt{5}} |\cB(\theta_1, x)\rangle = \sum_{(n_1, n_2, \tilde{n}_1, 0) \in S_0} \sum_{J=0}^\infty e^{i \tilde{n}_1 \theta_1} P_J(x) \Vert (n_1, n_2, \tilde{n}_1, 0); J \rangle\!\rangle \,.
\ee
where $-1 \leq x \leq 1$ is a parameter and $P_\ell(x)$ is a Legendre polynomial. However, as shown in \cite{Janik:2001hb}, these states always exhibit a continuous spectrum in the open channel, and so are somewhat pathological. We learn that the complete set of boundary states consists of $|\cB_\pm(\theta_1, \theta_2)\rangle$ and $|\cB(\theta_1, x)\rangle$, with only the former being physically reasonable, and these preserve an additional $U(1)'$ or $U(1)''$, as assumed.

\paragraph{Scaling dimensions in the 3450 Model.} Here we show that the scattering process~\eqref{eq:out_state_345} conserves scaling dimension:
\begin{equation}
n_1 \psi_1+n_2\psi_2\longrightarrow T^{n_1+3n_2}+\tilde n_1\tilde \psi_1+\tilde n_2\tilde \psi_2
\end{equation}
where
\begin{equation}
\begin{aligned}
\tilde n_1&=\tfrac35n_1+\tfrac45n_2-\tfrac12+\bigl[\tfrac12+\tfrac25n_1+\tfrac15n_2\bigr]\\
\tilde n_2&=\tfrac45n_1-\tfrac35n_2-\tfrac12+\bigl[\tfrac12+\tfrac15n_1-\tfrac25n_2\bigr]
\end{aligned}
\end{equation}

The in-state carries dimension $(\frac12(n_1^2+n_2^2),0)$, and therefore we expect the out-state to carry dimension $(0,\frac12(n_1^2+n_2^2))$. In order to check that this is indeed correct, we proceed as in section~\ref{sec:345}, that is, we study the twisted Hilbert space defined by the boundary conditions $\psi(\sigma+2\pi)=e^{2\pi i\eta}\psi(\sigma)$ (see figure~\ref{fig:twist_H}). Each fermion contributes a vacuum dimension of
\begin{equation}
\tfrac{c}{24}+\sum_{p\in\mathbb N+[\eta]}p=\tfrac{c}{24}+\sum_{p\in\mathbb N+[\eta]}p^{-s}\biggr|_{s\to-1}=\tfrac{c}{24}+\zeta(s,[\eta])\biggr|_{s\to-1}=\tfrac{1}{8}-\tfrac12[\eta](1-[\eta])
\end{equation}
where in the last equality we used $c=1$ for each complex fermion. Then, $|T^k\rangle$ carries $h_\text{vac}=0$ and 
\begin{equation}\label{eq:app_vac_h}
\tilde h_\text{vac}=\sum_\text{right}\tfrac{1}{8}-\tfrac12[\tilde \eta](1-[\tilde \eta])
\end{equation}

We also need to compute the scaling dimension of the oscillators $\tilde\psi^{\tilde n}$. Recall that this composite operator is defined via a subtracted OPE, cf.
\begin{equation}
\tilde\psi^{\tilde n}:=\prod_{j=0}^{\tilde n-1}\tilde\partial^j\tilde\psi
\end{equation}
Each derivative carries dimension $\tilde h(\tilde\partial^j)=j$, while the fermion $\tilde\psi$ itself carries dimension $\tilde h(\tilde\psi)=[\sign(\tilde n)\tilde\eta]$ (recall that negative $\tilde n$ means that we use $\tilde\psi^\dagger$). Thus, all in all, the oscillators carry scaling dimension
\begin{equation}\label{eq:app_psi_h}
\tilde h(\tilde\psi^{\tilde n})=\sum_{j=0}^{|\tilde n|-1}[\sign(\tilde n)\tilde\eta]+j\equiv \tfrac12\tilde n^2+\tilde n([\tilde\eta]-\tfrac12)
\end{equation}

Finally, using~\eqref{eq:app_vac_h} and~\eqref{eq:app_psi_h} (with $\tilde\eta_i=\frac12+\frac{q_ik}{5}$ with $q_i=3,4$ for $\tilde\psi_1,\tilde\psi_2$, respectively) one can easily check that
\begin{equation}
\tilde h_\text{vac}+\tilde h(\tilde\psi_1^{\tilde n_1})+\tilde h(\tilde\psi_2^{\tilde n_2})\equiv\frac12(n_1^2+n_2^2)
\end{equation}
and scaling dimension is conserved, as required.

More generally, if we have $N$ fermions charged under $U(1)^N$, then the general scattering experiment is fixed by charge conservation to $n_i X_i\longrightarrow n_i\mathcal R_{ij}\tilde X_j$. The condition that scaling dimension is preserved amounts to $||n||^2=||\mathcal Rn||^2$, which holds if and only if $\mathcal R$ is orthogonal, i.e., if $U(1)^N$ is anomaly-free. In the fermionic language, the conservation of scaling dimension is less manifest but can be checked using the same method as before.

\begin{center}***\end{center}

Let us also check the conservation of $\Z_5$ charge in the process~\eqref{eq:out_state_345}. Recall that the fermions have charges $q_i=0,0,3,4$, so the in-state is manifestly neutral. The $\Z_5$ quantum number of the out-state is the sum $Q_\text{vac}(T^{n_1+3n_2})+3 \tilde n_1+4 \tilde n_2$. Since $\tilde n_1, \tilde n_2$ are given in~\eqref{eq:out_tilden_1}, we need only compute the vacuum charge that the endpoint of the line adds:
\begin{equation}
\ba
Q_\text{vac}(T^{n_1+3n_2})&=\sum_i q_i\times\bigl(\tfrac12-\bigl[\tfrac12+\tfrac15q_i(n_1+3n_2)\bigr]\bigr)\gamma_i^\star\\
&=5(n_1+3n_2)-3 \left\lfloor \tfrac{1}{2}+\tfrac{3}{5} (n_1+3 n_2)\right\rfloor -4 \left\lfloor \tfrac{1}{2}+ \tfrac{4}{5} (n_1+3 n_2)\right\rfloor \,.
\ea
\end{equation}
Adding the contribution from the local excitations using~\eqref{eq:out_tilden_1} some straightforward algebra shows that
\be 
Q_\text{vac}(T^{n_1+3n_2})+3 \tilde n_1+4 \tilde n_2=5n_1 \equiv 0 \mod5\,.
\ee 
We have therefore verified by explicit computation that the scattering~\eqref{eq:out_state_345} preserves $\Z_5$ charge, as indeed it should since $\Z_5$ sits inside the $U(1) \times U(1)'$ symmetry preserved by the boundary state.

\paragraph{Monopole Scattering.} We now show that the process~\eqref{eq:monopole_higher_n_fermonic_out},
\begin{equation}\label{eq:app_higher_n}
\psi_{i,\mu}\longrightarrow T^{-2}+\tilde\psi_{i,\mu}
\end{equation}
conserves the $\mathbb Z_{nN_f}$ symmetry that acts as $\psi\mapsto\psi,\tilde\psi\mapsto e^{2\pi i/nN_f}\tilde\psi$. In the bosonic picture,
\begin{equation}\label{eq:app_out-state_n}
e^{i X_{i,\mu}}\longrightarrow e^{i\tilde X_{i,\mu}-\frac{2}{nN_f}i\sum_{j,\mu'}\tilde X_{j,\mu'}}
\end{equation}
and conservation of the symmetry amounts to checking that the out-state has no self-monodromy, which is straightforward.

However, in translating this to the fermionic picture, we must take into account that there is an ambiguity in the definition of the $\Z_{nN_f}$ charge $q$. Clearly, shifting $q \mapsto q+nN_f$ leads to the same transformation for integer vertex operators, but it does not in general for fractional ones. In particular, the transformation 
\be 
e^{-\tfrac{2}{nN_f}i\tilde X} \mapsto e^{-\tfrac{2}{nN_f}i\tilde X-\tfrac{4\pi i q}{(nN_f)^2}}
\ee
leads to different results for $q=1$ and $q=1+nN_f$. In this way $T^{-2}$ can be said to act by $\tilde X_i \mapsto \tilde X_i +\tfrac{2\pi q_i}{nN_f}$ with $q_i=-2$ or $q_i=nN_f-2$. In the outgoing state~\eqref{eq:app_out-state_n}, we take the charges to be $nN_f-2$ for $\tilde X_{i,\mu}$ and $-2$ for the rest. It is simple to check that this assignment of charges leaves the out-state invariant. Furthermore, using~\eqref{eq:general_Qvac}, one can check that this assignment leads to $Q_\text{vac}(T^{-2})=-1$. Hence, the out-state has overall charge $Q(T^{-2}+\tilde\psi)=-1+1\equiv0$, i.e., it is $\mathbb Z_{nN_f}$-neutral, just like the in-state.

It is important to stress that the assignment of charges being $nN_f-2$ for $\tilde X_{i,\mu}$ and $-2$ for the rest means that the line does not act the same on all fractional vertex operators $e^{i\alpha\tilde X}$, but it still acts as $\tilde \psi \mapsto e^{2\pi i/nN_f}\tilde \psi$ on the local fermions. This is an artifact of using abelian bosonization.

\begin{center}***\end{center}

Let us also check that~\eqref{eq:app_higher_n} conserves scaling dimension. In the bosonic picture~\eqref{eq:app_out-state_n} this is manifest, but it is less obvious in the fermionic picture. Using~\eqref{eq:app_vac_h} one can check that the endpoint of $T^{-2}$ has weight $\tilde h_\text{vac}=\frac{2}{nN_f}$ while the fermion itself has scaling dimension $\tilde h(\tilde \psi)=\frac{1}{2}-\frac{2}{nN_f}$. Hence, the scaling dimension of the out-state $T^{-2}+\tilde\psi$ is $\tilde h\equiv\frac12$, matching the dimension of the in-state.

\section{Review of some $\boldsymbol{2d}$ Anomalies} 
\label{app:anomalies}

Here we collect some known facts about anomalies in $2d$. We always assume that $c_L=c_R$ so that there are no purely gravitational anomalies.

First, given a pair of $U(1)$ symmetries, the perturbative $U(1)$-$U(1)'$ anomaly is
\begin{equation}
\sum_i Q_i Q_i'\gamma_i^\star
\end{equation}
where $\gamma^\star=\pm1$ is the chirality matrix, and the sum runs over all fermions in the theory.

Discrete symmetries are a little bit trickier. This is illustrated by the fact that~\cite{Cheng_2018,Grigoletto:2021zyv}
\begin{equation}
\Omega_3^\text{spin}(B\mathbb Z_n)=\begin{cases}
\mathbb Z_n & n\text{ odd}\\
\mathbb Z_2\times\mathbb Z_{2n}&n/2\text{ even}\\
\mathbb Z_{4n}&n/2\text{ odd}
\end{cases}
\end{equation}
which classifies anomalies for $\mathbb Z_n$ symmetries.

For $n$ odd, all anomalies are induced from the embedding $\mathbb Z_n\subset U(1)$, namely, they are the mod $n$ reduction of the perturbative anomaly above. For example, the $\mathbb Z_n$-$\mathbb Z_n$ anomaly is
\begin{equation}\label{eq:ap_n_odd}
\sum_i q^2_i\gamma_i^\star\mod n\qquad\text{(for $n$ odd)}
\end{equation}
where $q_i=0,1,\dots,n-1$ denotes the charge of the fermions (in the sense that, under $k\in\mathbb Z_n$, they transform as $\psi_i\mapsto e^{2\pi i q_ik/n}\psi_i$).

For $n$ even, things are not quite so simple. The mixed $\mathbb Z_n$-$U(1)$ anomaly is still given by the mod $n$ reduction of the perturbative anomaly, but the $\mathbb Z_n$-$\mathbb Z_n$ anomaly requires more work. In this case, the induced anomaly is still meaningful, but it only detects the $\mathbb Z_n$ subgroup of $\mathbb Z_{2n}$ or $\mathbb Z_{4n}$.
 It turns out that, for even $n$, the answer is
 \begin{equation}\label{eq:ap_n_even}
\sum_i q^2_i\gamma_i^\star\mod 2n\qquad\text{(for $n$ even)}
\end{equation}
as we show below.

In order to detect the full non-perturbative anomaly, a standard procedure is to look at the spin of the operators in the $\mathbb Z_n$ twisted Hilbert space~\cite{Chang:2018iay,Lin:2019kpn,Delmastro:2021xox}.\footnote{Of course, here we implicitly assume that such twisted Hilbert space actually exists in the first place. This is indeed the case when all fermions are Dirac, but an odd number of Majorana-Weyl fermions do not admit a well-defined, $\mathbb Z_2$-graded Hilbert space.} Let us focus on the Ramond twisted Hilbert space, defined by
\begin{equation}
\psi_i(\sigma+2\pi)=e^{2\pi i q_i /n}\psi_i(\sigma)
\end{equation}
In an anomaly-free theory, an $n$-fold rotation should act trivially on this Hilbert space. (We look at the R-sector for simplicity but if we were to look at the NS sector, the claim would be that the $n$-fold rotation should act as $(-1)^n$). An anomaly for $\mathbb Z_n$ is then detected by exhibiting an operator whose spin is not an integral multiple of $1/n$, but rather has a larger denominator. In other words, given that all creation operators have integral conformal weight, the anomaly of $\mathbb Z_n$ is measured by $nE^\text{R}_0\mod 1$, where $E_0^\text{R}$ is the vacuum energy of the twisted Ramond Hilbert space. (As before, if we were to look at the NS sector, the anomaly would be $n(E_0^\text{NS}+1/2)\mod1$.)

This vacuum energy is easy to compute. Each mode with boundary condition $\psi\mapsto e^{2\pi i \eta}\psi$ contributes
\begin{equation}
\sum_{p\in\mathbb N+[\eta]}p=\sum_{p\in\mathbb N+[\eta]}p^{-s}\biggr|_{s\to-1}=\zeta(s,1-[\eta])\biggr|_{s\to-1}\equiv-\frac{1}{12}+\frac12[\eta](1-[\eta])
\end{equation}
and therefore the total vacuum energy is
\begin{equation}
E^\text{R}_0=\sum_i \bigl(-\tfrac{1}{12}+\tfrac12\bigl[q_i/n\bigr]\bigl(1-\bigl[q_i/n\bigr]\bigr)\bigr)\gamma_i^\star
\end{equation}

Given that $[\eta]=\eta+\text{integer}$ and $c_L=c_R$, this expression can be simplified into
\begin{equation}
nE^\text{R}_0=\frac12\sum_i q_i(1-q_i/n)\gamma_i^\star\mod1
\end{equation}
The fractional part of this expression then gives the anomaly of $\mathbb Z_n$. It is now a simple exercise to show that
\begin{equation}
\frac12\sum_i q_i(1-q_i/n)\gamma_i^\star\equiv0\mod1\qquad\Longleftrightarrow\qquad \sum_i q_i^2\gamma_i^\star\equiv0\mod \gcd(n,2)n
\end{equation}
in agreement with~\eqref{eq:ap_n_odd} and~\eqref{eq:ap_n_even}.

Let us check a few examples. First off, if both left-movers and right-movers transform with the same charges, $q_i=\tilde q_i$, the anomaly manifestly vanishes, even the non-perturbative part. We could in fact have guessed this since in this case it is possible to write down a $\mathbb Z_n$-preserving mass term.
Likewise, if they transform with opposite charges, $q_i=-\tilde q_i$, the anomaly is $nE_0^\text{R}=\sum_i q_i\mod1$, which vanishes for integer charges.

For a more interesting example, consider $N$ fermions and a $\mathbb Z_2$ symmetry where all left-movers transform as $\psi\to-\psi$ and all right-movers as $\psi\to+\psi$. Then, the anomaly is
\begin{equation}
2E^\text{R}_0=\frac12N(1-1/2)\equiv \frac14N\mod 1
\end{equation}
which agrees with the known answer~\cite{Fidkowski_2010,Ryu_2012,Gu:2013azn} (the general result is that $\nu$ Majorana fermions have anomaly $\nu/8\mod1$, and our fermions are Dirac, so $N=2\nu$).

Finally, take $N$ fermions and a $\mathbb Z_N$ symmetry where all right-movers transform with charge $+1$, and all left-movers are neutral. The anomaly reads
\begin{equation}
NE^\text{R}_0=-\frac12N(1-1/N)\equiv \frac12(1-N)\mod 1
\end{equation}
This vanishes if $N$ is odd, in agreement with the fact that the perturbative anomaly vanishes, but it is non-zero for $N$ even, showing that for $N$ even the symmetry is anomalous. On the other hand, for $N$ even, the $\mathbb Z_{N/2}$ subgroup is anomaly-free (to show this we just replace $1/N\to 2/N$ in the formula above, with result $(N/2)E_0^\text{R}=(2-N)/2$, which indeed vanishes mod $1$). We then learn that only the $\mathbb Z_2\subset\mathbb Z_N$ subgroup suffers from non-perturbative anomalies, while the anomaly of $\mathbb Z_{N/2}$ is determined by the usual perturbative formula.

\section{Lagrangian for Hall state}\label{app:hall_TQFT}

In this appendix we collect some properties of the Hall state TQFT $\mathcal A^{N,p}$ that we use throughout the text. This TQFT is labeled by a pair of integers $N,p$ such that $\gcd(N,p)=1$~\cite{Hsin:2018vcg}. By definition, the fusion algebra is group-like and isomorphic to $\mathbb Z_N$; and, furthermore, the generator of $\mathbb Z_N$ carries topological spin $\frac{p}{2N}\mod\frac12$. Equivalently, $\mathcal A^{N,p}$ is an abelian TQFT with $1$-form symmetry $\mathbb Z_N^{(1)}$ and anomaly $p/N\mod 1$. Note that we regard $\mathcal A^{N,p}$ as a spin TQFT. Moreover, it is often convenient to stack the theory with some invertible TQFT such that the total central charge vanishes; this can be accomplished by adding factors of $U(1)_{\pm1}$ as necessary (which we always leave implicit). When the central charge vanishes, the defect $\mathcal D_{p/N}$ becomes truly topological; otherwise, its phase depends on the choice of trivialization of the tangent bundle of its support~\cite{Putrov:2023jqi}.

Abstractly, the anyons of the Hall state are labeled by an integer $s\in\{0,1,\dots,N-1\}$ and they satisfy fusion rules $s\times s'=(s+s'\mod N)$. They carry spin $h_s=\frac{ps^2}{2N}$ and braid with the generator $s=1$ with phase $e^{-2\pi i sp/N}$. For some purposes, it may be useful to have an explicit Lagrangian for this TQFT. Being abelian, such theory can always be written as a $U(1)^q$ Chern-Simons theory for some $q$, namely
\begin{equation}
\mathcal L=\frac{1}{4\pi}a^t K \mathrm da+\frac{1}{2\pi}a^t v\,\mathrm dA
\end{equation}
where $K$ is a $q\times q$ integral symmetric matrix, and $v$ is a $q\times 1$ vector of integers. On the other hand, $A$ is the electromagnetic field (thought of as a background field here).

The conductivity $p/N$ can be obtained by formally integrating out $a$ using its equations of motion
\begin{equation}\label{eq:Hall_EOM}
Ka+Av=0
\end{equation}
and plugging the result back to $\mathcal L$, cf.
\begin{equation}
\mathcal L=-\frac{1}{2\pi}(v^tK^{-1}v)A\mathrm dA
\end{equation}
from where we read the conductivity as $p/N=v^t K^{-1}v$. The line that generates $\mathbb Z_N^{(1)}$ can be taken as $W=\exp\bigl(iv^t\int a\bigr)$, and an arbitrary line can be written as $W^s$. The spin of $W^s$ is $\frac12 (sv)^tK^{-1}(sv)\equiv \frac{ps^2}{2N}$, as above. Using~\eqref{eq:Hall_EOM}, it is easily checked that $W^s$ couples to $A$ with coefficient $sv^tK^{-1}v$, namely this anyon carries electromagnetic charge $sp/N$.\footnote{Note that the lines $s=0,1,\dots,N-1$ all carry fractional charge, and only the transparent anyon $W^N$ has integral charge. Indeed, transparent anyons describe the worldlines of microscopic fermions.}

Although the explicit form of $K$ is seldom needed, it may be useful to know that a general expression is known (see e.g.~\cite{Tong:2016kpv}). For example, a state with conductivity $p/N$ can be constructed by taking the $K$-matrix
\begin{equation}
K=\begin{pmatrix}
+k_1&1&0&0&0&\cdots\\
1&-k_2&1&0&0&\cdots\\
0&1&+k_3&1&0&\cdots\\
0&0&1&-k_4&1&\cdots\\
\vdots&\vdots&\vdots&\vdots&\vdots&\ddots
\end{pmatrix}
\end{equation}
where $k_i$ are the coefficients of the continued fraction of $p/N$, namely
\begin{equation}
\frac{p}{N}=\frac{1}{k_1+\frac{1}{k_2+\frac{1}{k_3+\cdots}}}
\end{equation}
The charge vector is $v=(1,0,\dots)$.

For example, if $K=(N)$ is $1\times 1$ and $v=1$, we find a fractional coefficient $1/N$, which means that $U(1)_N$ realizes the theory $\mathcal A^{N,1}$. On the other hand, if we take $N$ to be an odd integer and take
\begin{equation}\label{eq:TQFT_n_odd}
K=\begin{pmatrix}
\frac{N+1}{2}&1\\1&2
\end{pmatrix}
\end{equation}
and $v=(1\ 0)$, then the fractional coefficient becomes $2/N$, which means this $K$ matrix realizes $\mathcal A^{N,2}$. As a quick check, this matrix has Smith normal form
\begin{equation}
K=\begin{pmatrix}
\frac{N+1}{2}&-1\\1&0
\end{pmatrix}\begin{pmatrix}
1&0\\0&N
\end{pmatrix}\begin{pmatrix}
1&2\\0&1
\end{pmatrix}
\end{equation}
which means that the $1$-form symmetry is $\mathbb Z_N^{(1)}$. Also, a generic line can be written as $W=(s,0)$, whose spin is $h_s=s^2/N$, which indeed confirms that this has $p=2$.

In conclusion, the Hall state for the topological defect $\mathcal D_{2/n}$ that we use in the main text admits a Lagrangian description of the form
\begin{equation}
\begin{cases}
K=\begin{pmatrix}n/2\end{pmatrix}&n\text{ even}\\[+2ex]
K=\begin{pmatrix}
\frac{n+1}{2}&1\\1&2
\end{pmatrix}&n\text{ odd}
\end{cases}
\end{equation}

\clearpage
\printbibliography
\end{document}